\documentclass[10.5pt,a4paper]{article}

\usepackage[T1]{fontenc}
\usepackage{lmodern}
\usepackage{amsmath,amssymb,amsthm}
\usepackage{geometry}
\usepackage{graphicx}
\usepackage{booktabs}
\usepackage{array}
\usepackage{multirow}
\usepackage{float}
\usepackage{authblk}

\usepackage{microtype}
\usepackage{indentfirst}
\usepackage{needspace}
\usepackage{xcolor}
\usepackage{tikz}
\usepackage{algorithm}
\usepackage{algpseudocode}
\usepackage[numbers,sort&compress]{natbib}
\usepackage{hyperref}

\makeatletter
\newenvironment{breakablealgorithm}[1]{%
  \par\medskip
  \noindent\hrule height0.8pt depth0pt\kern2pt
  \refstepcounter{algorithm}\phantomsection
  \noindent\textbf{\fname@algorithm~\thealgorithm}\hspace{0.6em}#1\par
  \nopagebreak\kern2pt\hrule\kern2pt
  \small
  \ignorespaces
}{%
  \nopagebreak\kern2pt\hrule\relax
  \par\medskip
}
\makeatother

\newtheorem{theorem}{Theorem}
\newtheorem{lemma}{Lemma}
\newtheorem{corollary}{Corollary}

\newcommand{\R}{\mathbb R}
\newcommand{\Z}{\mathbb Z}
\newcommand{\Ctilde}{\widetilde{\mathbf C}}
\newcommand{\Nper}{\mathbf N_{\mathrm{per}}}

\newlength{\parindentthree}
\hypersetup{
  colorlinks=true,
  linkcolor=blue,
  citecolor=blue,
  urlcolor=blue,
  pdftitle={Fairness-Optimized Interlacing Knots for Even-Degree Periodic B-Spline Interpolation},
  pdfsubject={Interlacing domain knots and fairness optimization for periodic B-spline interpolation},
  pdfkeywords={periodic B-spline, interpolation, even degree, interlacing knots, fairness energy, optimization}
}

\title{\large \bf Fairness-optimized interlacing knots for even-degree periodic B-spline interpolation}
\author[a,*]{Shutao Tang}
\author[b]{Zixuan Zhao}
\affil[a]{\small Department of Mechanical Engineering, Northwestern Polytechnical University, Xi'an, China}
\affil[b]{\small The University of Chicago, 5801 S Ellis Ave, Chicago, IL 60637, United States}
\date{}

\begin{document}
\maketitle
\vspace{-1.2em}

\begin{abstract}
\noindent Interpolating an ordered set of points by a B-spline curve is a
fundamental operation in computer-aided geometric design. Most of the literature
addresses open curves, where the Schoenberg--Whitney condition characterizes a
nonsingular collocation matrix and averaging is the standard way to choose the
knot vector. Periodic B-spline interpolation has received much less attention, and
the open-curve algorithms cannot be transferred directly. No theoretical criterion
is available for choosing the knots, and the few existing constructions, such as
the shifting and energy methods, remain empirical, with neither a theoretical
guarantee nor a supporting analysis. This paper addresses this gap for even degrees.
We prove that for every even degree $p\ge2$ the periodic collocation matrix is
nonsingular whenever the domain knots and the interpolation parameters interlace,
in either of two admissible orderings, for arbitrarily nonuniform parameters.
Rather than singling out one knot vector, this sufficient condition delimits an
$n$-parameter family of admissible knot vectors and provides the theoretical
foundation for knot selection that periodic interpolation has lacked. It also
explains why the empirical shifting method is solvable, previously justified only
by an incorrect diagonal-dominance argument. We then exploit the remaining freedom
to place the knots by fairness optimization, which preserves admissibility at every
iterate and leads to the methods Opt-R and Opt-L. On four closed shapes with
irregularly distributed points these methods achieve a fairness comparable to the
shifting and energy methods, improve on them in several cases, and offer guarantees
that no empirical construction provides.

\vspace{0.5em}

\end{abstract}

\section{Introduction}
\label{sec:introduction}

Interpolating an ordered set of points with a B-spline curve is one of the
most common operations in computer-aided geometric design (CAGD), and it requires
two preliminary decisions: a parameter value, or node, has to be assigned to
every data point, and a knot vector has to be chosen for the spline. These
two choices determine the collocation matrix that relates the control points to
the data, and hence both the shape of the resulting curve and the numerical
behaviour of the interpolation system. For open-curve interpolation,
much of the classical literature is devoted to the choice of the parameters,
and once they are available the knot vector is usually obtained by the classical
averaging technique \cite{de1978practical,piegl1996nurbs}, which yields a
totally positive banded collocation matrix that can be solved by Gaussian
elimination without pivoting. For closed, or periodic, B-spline curves the situation is
more delicate, because the cyclic structure of the curve introduces additional linear
constraints and the averaging technique cannot be applied directly. The extensive
literature has concentrated on open curves, and these open-curve techniques cannot be
transferred directly to the periodic setting, which has received comparatively little
attention.

For open-curve interpolation the widely used Schoenberg--Whitney condition
\cite{de1978practical} characterizes the parameters and knots for which the
collocation matrix is nonsingular. However, no comparable condition is known
for periodic B-spline curves, which have been approached mainly through
empirical constructions tailored to particular degrees
\cite{park1996smooth,park2001choosing,park2003lofted}, and studies of periodic
B-spline interpolation are comparatively few \cite{tang2023control}. No simple
characterization of the knot vectors for which the periodic interpolation
system is solvable is available, and there is a lack of constructive rules that
guarantee a well-conditioned interpolation system for periodic
B-spline curves. The method proposed in this paper is intended to fill this gap.

This paper concerns the even-degree case and is built on a simple observation.
For an even degree $p$, the periodic collocation matrix is nonsingular whenever
the domain knots and the interpolation parameters interlace, in the sense
that the initial knot coincides with the first parameter and each interior
knot lies strictly between two consecutive parameters (Figure~\ref{fig:interlacing} or \ref{fig:left-interlacing}). The parameters may be
arbitrarily nonuniform, and every interior knot may be placed independently
inside its interval, so that the interlacing condition describes not a single
knot vector but a whole family of admissible ones. It is the periodic,
even-degree counterpart of the Schoenberg--Whitney condition for open curves.
In particular, because the shifted knots of the \emph{shifting} method satisfy
the left interlacing ordering, that method is guaranteed to be solvable, a fact
previously justified only by an incorrect diagonal-dominance argument~\cite{park2001choosing}. Unique
solvability is thus guaranteed for a substantial set of knot vectors,
and the remaining freedom can be used to improve the quality of the
interpolating curve.

We use this freedom to place the knots by fairness optimization. Each
interior knot is placed inside its admissible interval by one coefficient
$\theta_i\in(0,1)$, and the coefficients are computed by locally minimizing
a two-term fairness energy combining the first and second derivative
energies of the resulting curve. The box constraint on the coefficients is
removed by a logistic reparametrization, and the resulting unconstrained
problem is solved numerically by a multi-start quasi-Newton method. Because
the coefficients remain in the open unit interval at every iterate, the
interlacing condition, and hence unique solvability, holds throughout the
optimization. The resulting methods Opt-R and Opt-L are compared with the
two deterministic even-degree constructions, the \emph{shifting} method and the
\emph{energy} method \cite{park2001choosing}, on four closed shapes with
complex, nonuniform point distributions, using three parametrizations:
chord-length, centripetal \cite{lee1989choosing}, and the improved
parametrization of Fang and Hung \cite{fang2013improved}.

The main contributions of this work can be summarized as follows:

\begin{itemize}
  \item \textbf{Interlacing condition for unique solvability.}
        We prove that for every even degree $p\ge2$ the
        periodic collocation matrix is nonsingular whenever the domain knots
        interlace with the parameters, in either the right or the left
        ordering. The parameters may be arbitrarily nonuniform, and the
        interior knots may be chosen independently within their intervals
        (Theorem~\ref{thm:main} and Corollary~\ref{cor:left}).

  \item \textbf{Consequences for existing constructions.}
        The interlacing condition explains why the empirical \emph{shifting}
        method is solvable, despite the failure of Park's diagonal-dominance
        argument \cite{park2001choosing}.

  \item \textbf{Fairness-optimized knot placement.}
        We turn the $n$-parameter family of admissible knot vectors left by
        the interlacing condition into a fairness problem and treat it with a
        logistic reparametrization and a multi-start quasi-Newton local search that
        preserves admissibility at every iterate, yielding the two locally optimized
        constructions Opt-R and Opt-L (Section~\ref{sec:fairness}).

  \item \textbf{Comparison with deterministic even-degree constructions.}
        We compare Opt-R and Opt-L with the \emph{shifting}
        and \emph{energy} methods using three parametrizations (chord-length,
        centripetal~\cite{lee1989choosing}, and the improved parametrization of
        Fang and Hung \cite{fang2013improved}), and identify the trade-offs between
        fairness, conditioning, and computational cost
        (Section~\ref{sec:experiments}).
\end{itemize}

Table~\ref{tab:comparison} contrasts the proposed approach with the standard
knot-selection strategies and highlights the theoretical guarantee that the
existing constructions lack.

\begin{table}[htb]
\centering
\caption{Comparison of knot-selection strategies for B-spline curve interpolation.}
\label{tab:comparison}
\small
\setlength{\tabcolsep}{4pt}
\renewcommand{\arraystretch}{1.35}
\begin{tabular}{@{}>{\raggedright\arraybackslash}p{0.14\textwidth}>{\raggedright\arraybackslash}p{0.085\textwidth}>{\raggedright\arraybackslash}p{0.19\textwidth}>{\raggedright\arraybackslash}p{0.26\textwidth}>{\raggedright\arraybackslash}p{0.255\textwidth}@{}}
\toprule
Method & Category & Strategy & Advantages & Disadvantages\\
\midrule
Averaging~\cite{de1978practical,piegl1996nurbs}
  & Open
  & Average of $p$ consecutive parameters.
  & Standard for open curves; banded totally positive system.
  & Not applicable to periodic curves; ill-conditioned or singular for even degree.\\[2pt]
Shifting~\cite{park2001choosing}
  & Closed
  & Half-spans of adjacent parameters; shifts the parameters.
  & Simple; direct construction.
  & Empirical; no theoretical proof or conditioning guarantee.\\[2pt]
Energy~\cite{park2001choosing,park2003lofted}
  & Closed
  & Extra control point; constrained energy minimization.
  & Low energy; smooth curves.
  & Empirical; no theoretical guarantee that the collocation matrix has full rank.\\[2pt]
Opt-R, Opt-L\newline (this paper)
  & Closed
  & Interlaced knots; fairness-energy minimization.
  & Solvability guaranteed at any even degree; knot freedom; proves shifting solvable.
  & Sufficient only; even degree; multi-start cost.\\
\bottomrule
\end{tabular}
\end{table}

The rest of the paper is organized as follows.
Section~\ref{sec:related-work} reviews the related work on B-spline curve
interpolation. Section~\ref{sec:preliminaries} introduces the technical
background and states the problem addressed in the paper.
Section~\ref{sec:interlacing} proves the interlacing condition for unique
solvability. Section~\ref{sec:fairness} formulates the
fairness-optimized knot placement problem and its local numerical solution.
Section~\ref{sec:experiments}
presents experimental results, and Section~\ref{sec:conclusions} concludes
the paper and outlines future work.

\section{Related work}
\label{sec:related-work}

This section reviews the main lines of work related to B-spline curve
interpolation: the parametrization of the data, the determination of the knot
vector, the optimization strategies that treat parameters and knots as
unknowns, and the numerical conditions that guarantee a well-posed
interpolation.

\subsection{Data parametrization}
\label{subsec:parametrization}

Parametrization maps each data point to a value in the parameter domain, and the
quality of an interpolation curve depends heavily on this mapping. Several
classical strategies are in common use. The uniform method distributes the
parameters evenly, the chord length method~\cite{piegl1996nurbs} 
distributes them in proportion to the distance between consecutive points, and the centripetal
method of Lee \cite{lee1989choosing} uses the square root of that distance. Foley
and Nielson \cite{foley1989knot} take the angle between neighbouring segments into
account so as to obtain a parametrization that is invariant under affine
transformations, whereas Lim \cite{lim1999universal} and Jung and Kim
\cite{jung2000new} place a node where the corresponding B-spline basis function
attains its maximum. None of these heuristics is universally satisfactory: uniform
parameters can generate loops or oscillations when the points are unevenly spaced,
chord length parameters are sensitive to abrupt changes of direction, while
centripetal parameters are more robust in such cases. Subsequent work has refined
the centripetal idea to suppress undesirable wiggle \cite{fang2013improved}, or has
combined the geometric and basis-oriented viewpoints in hybrid schemes
\cite{haron2012parameterization}. Dynamic and modified chord length variants
\cite{balta2020dynamic,xu2022improved}, as well as node rules derived from local
cubic polynomials \cite{wang2024computing}, follow the same heuristic philosophy.

More recently, data-driven approaches have been explored. Laube et al.
\cite{laube2018deep} train a deep network to predict the parameters and knots of an
approximating curve, Scholz and J{\"u}ttler \cite{scholz2021parameterization} use a
residual network to parametrize polynomial curves, and Song et al.
\cite{song2024hybrid} learn a hybrid parametrization by supervised learning, while
Lin and Feng \cite{lin2025parameterization} regress the parameters for B-spline
curve interpolation.
Tang and Yang \cite{tang2025dnn} propose a network that directly produces the
parametrization for B-spline approximation, and extend it to the joint prediction
of knots and parameters \cite{tang2026neural}; a recent review surveys this family
of methods \cite{zhu2023review}. Such methods adapt well to the statistics of the
training data, but their behaviour depends on the training set, generalization is
not guaranteed, and they provide no optimality statement for an individual
dataset.

\subsection{Knot determination with prescribed parameters}
\label{subsec:knots}

Once the parameters have been prescribed, the knot vector still has to be determined,
and different knots lead to different curves and to different conditioning of the
interpolation system. For an open curve the standard tool is the averaging
technique \cite{piegl1996nurbs,de1978practical}, in
which each interior knot is the average of $p$ consecutive parameters
(here $p$ is the degree of the spline); the
resulting system matrix is totally positive and banded with a semi-bandwidth
smaller than $p$, so that it can be solved by Gaussian elimination without
pivoting \cite{piegl1996nurbs,de1978practical}. The knots thereby reflect the
distribution of the parameters.

\begin{figure}[htb]
  \centering
  \resizebox{0.90\textwidth}{!}{%
  \begin{tikzpicture}[
      font=\small,
      branch/.style={draw=black!75, line width=0.6pt, rounded corners=3pt,
                     fill=white, font=\bfseries\small, inner xsep=7pt, inner ysep=5pt},
      pill/.style={draw=black!75, line width=0.6pt, rounded corners=9pt,
                   fill=white, font=\bfseries\footnotesize, inner xsep=7pt, inner ysep=4pt},
      link/.style={draw=blue!70!black, line width=0.9pt},
      leaf/.style={anchor=west, font=\small, text=black},
    ]

    \def\xa{-2.65}\def\xb{2.65}\def\ya{-0.8}\def\yb{0.8}\def\fc{0.46}
    \draw[black!75, line width=0.6pt, fill=blue!4]
        (\xa,\ya) -- (\xa,\yb) -- (\xb-\fc,\yb) -- (\xb,\yb-\fc) -- (\xb,\ya) -- cycle;
    \draw[black!75, line width=0.6pt, fill=blue!12]
        (\xb-\fc,\yb) -- (\xb-\fc,\yb-\fc) -- (\xb,\yb-\fc);
    \node[text=blue!55!black, font=\bfseries\small, align=center]
        at (0,0) {knot determination for\\B-spline curve interpolation};

    \node[branch] (open)   at (5.6, 1.75) {open};
    \node[branch] (closed) at (5.6,-1.75) {closed};
    \node[pill]   (odd)    at (9.4,-0.55) {odd-degree};
    \node[pill]   (even)   at (9.4,-2.95) {even-degree};

    \node[leaf] (l1) at (12.9, 1.75) {averaging technique \cite{piegl1996nurbs,de1978practical}};
    \node[leaf] (l2) at (12.9,-0.10) {averaging method \cite{park1996smooth}};
    \node[leaf] (l3) at (12.9,-1.15) {natural method \cite{de1978practical}};
    \node[leaf] (l4) at (12.9,-2.45) {shifting method \cite{park2001choosing}};
    \node[leaf] (l5) at (12.9,-3.55) {energy method \cite{park2001choosing}};

    \foreach \n in {l1,l2,l3,l4,l5}
        \draw[link] (\n.south west) -- (\n.south east);

    \draw[link] (\xb,0) to[out=22,in=180]  (open.west);
    \draw[link] (\xb,0) to[out=-22,in=180] (closed.west);
    \draw[link] (open.east)   to[out=0,in=180]   (l1.west);
    \draw[link] (closed.east) to[out=24,in=180]  (odd.west);
    \draw[link] (closed.east) to[out=-24,in=180] (even.west);
    \draw[link] (odd.east)    to[out=22,in=180]  (l2.west);
    \draw[link] (odd.east)    to[out=-18,in=180] (l3.west);
    \draw[link] (even.east)   to[out=22,in=180]  (l4.west);
    \draw[link] (even.east)   to[out=-18,in=180] (l5.west);
  \end{tikzpicture}%
  }
  \caption{Overview of knot determination methods for B-spline curve
  interpolation. For a periodic B-spline curve the available methods depend on the parity of
  the degree: averaging and natural methods are for odd degrees, while the
  shifting and energy methods were introduced for even degrees.}
  \label{fig:knot-summary}
\end{figure}
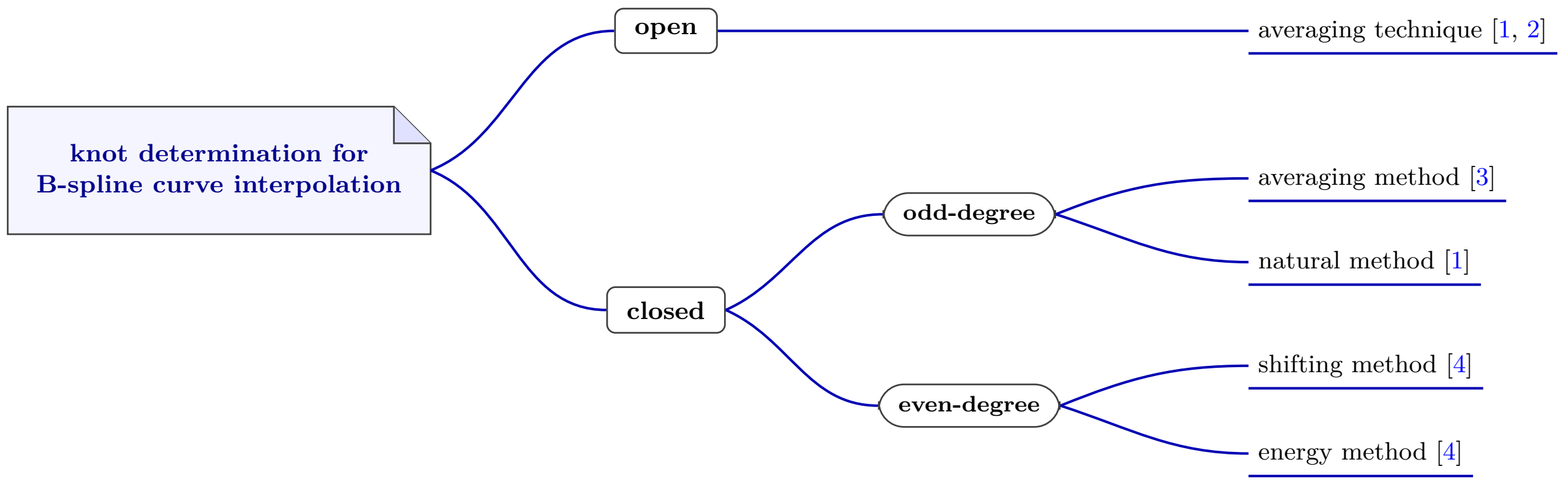

For periodic B-spline curve interpolation the averaging technique cannot be
applied directly, because the cyclic structure of the curve introduces additional
linear constraints. The situation depends on the parity of the degree. When the
degree is odd, averaging \cite{park1996smooth} and natural \cite{de1978practical}
strategies perform well. When the degree is even, however, averaging tends to
produce an ill-conditioned system matrix and may even make it singular
\cite{park2001choosing}. To overcome this difficulty, Park \cite{park2001choosing}
introduced the shifting and energy methods. The shifting method builds every knot
span from halves of two adjacent parameter spans and translates the parameters by
half of the last span, while the energy method increases the number of control
points by one, takes parameter midpoints as knots, and recovers the control points
by a strain energy minimization under linear constraints \cite{park2003lofted}.
Both produce visually pleasant curves, but the shifting method displaces the
parameters, so the start of the parameter domain is not the first data point.
Figure~\ref{fig:knot-summary} summarizes
the methods discussed above.

However, the shifting method is empirical; in particular, Park
\cite{park2001choosing} argues that simple pivoting makes
its collocation matrix strictly diagonally dominant. 
This claim is false; Appendix~\ref{app:park-dominance} gives a counterexample. 
This does not prevent the shifting method from being solvable: each
shifted domain knot is the midpoint of two consecutive shifted parameters, so
the shifted knots satisfy the left interlacing condition
\eqref{eq:left-interlacing}, and Corollary~\ref{cor:left} guarantees that the
shifting system is nonsingular. The energy method is empirical as well: its
constrained minimum is characterized by the KKT system \eqref{eq:kkt}, whose
solvability requires the rectangular collocation matrix to have full row rank
\cite{wang2008reducing,benzi2005numerical}, a property that is not proved in
Park's work.

\subsection{Optimization-based approaches}
\label{subsec:optimization}

A parallel line of research also addresses B-spline curve interpolation, treating
the parameters, the knot vector, and the control points as unknowns of an
optimization problem; formulations developed for curve approximation are not
considered here. The parameters have also been optimized, for example by
Kouibia et al. \cite{kouibia2011optimization}, who optimize the parameters of
cubic spline interpolation. Since the knot vector controls both the shape and the
solvability of the interpolation system, other works optimize the knots directly.
Thakur
characterizes optimal interpolatory quadratic splines with free knots as a convex
program \cite{thakur1986computable} and later solves a linear-programming
relaxation for near-optimal free-knot splines \cite{thakur2021data}; Suchomski
\cite{suchomski1991optimal} minimizes a discrete $L_2$ error over variable knots;
and Idais et al. \cite{idais2019optimal} allocate the knots of cubic and bicubic
interpolants with a multi-objective genetic algorithm.

More recent methods optimize the degrees of freedom jointly. Zhang et al.
\cite{zhang2025shape} directly minimize a curve-quality functional over the
control points, the parameters, and the knot vector at the same time;
they reformulate the interpolation constraints as an unconstrained problem and
prove that the distance between the resulting curve and the data points is bounded.
Hu and Zhang \cite{hu2020nsga} search for nodes and knots with the NSGA-II
multi-objective evolutionary algorithm, constraining the control points and the
tangent directions. Because these problems remain high-dimensional and nonconvex, the optimization-based
methods are computationally expensive and prone to local minima. All of these
studies, however, concern open curves; the optimization-based treatment of
periodic B-spline curve interpolation has been largely neglected.

\subsection{Numerical stability conditions}
\label{subsec:stability}

For open B-spline curve interpolation, the classical Schoenberg--Whitney
theorem \cite{de1978practical} is the necessary and sufficient condition for
the interpolation matrix to be nonsingular: every data parameter must lie
strictly inside the support of the corresponding B-spline basis function. This
condition is the basis of the numerical stability of open-curve B-spline
interpolation.

The periodic setting is structurally different. Because the knot vector is cyclic
and the control points wrap around, the collocation matrix is no longer a finite
banded matrix on an interval but a cyclically banded, circulant-like matrix on a
circle, so the classical Schoenberg--Whitney theorem cannot be applied in its
original form. The natural periodic analogue replaces the support of a basis
function by a window of consecutive knots centred on each parameter: each
parameter must lie strictly inside a window whose size depends on the degree,
comprising a parity-dependent number of knots on either side of the parameter.
This \emph{periodic window condition} was proposed by Tang
\cite{tang2023control} as the periodic counterpart of the Schoenberg--Whitney
condition, with the claim that it guarantees a nonsingular square periodic
collocation matrix. However, the proposed condition is not valid:
Appendix~\ref{app:window-counterexample} gives a uniform-knot configuration that
satisfies it while the periodic collocation matrix is singular. This disproves
Tang's condition, and the exact characterization of the knots and parameters
for which the periodic collocation matrix is nonsingular remains an open
problem. The
interlacing condition studied in this paper is a sufficient condition of a
different, constructive type: it
prescribes a family of knots for which nonsingularity is guaranteed at every
even degree.

\section{Problem statement and preliminaries}
\label{sec:preliminaries}

For the reader's convenience, Table~\ref{tab:notation} collects the principal
symbols and the methods compared in this paper.

\begin{table}[!ht]
\centering
\caption{Notation, definitions, and the methods compared in this paper.}
\label{tab:notation}
\small
\begin{tabular}{ll}
\toprule
Symbol & Meaning\\
\midrule
$p$, $n$ & even degree; last data-point index ($n+1$ points)\\
$\mathbf Q_i$ & data points, $0\le i\le n$\\
$t_i$ & interpolation parameters, $0=t_0<\cdots<t_n<1$\\
$u_i$ & domain knots, $0=u_0<\cdots<u_{n+1}=1$\\
$\mathcal N_{j,p}$ & periodic B-spline basis functions, $0\le j\le n$\\
$\widetilde{\mathbf C}(u)$ & periodic B-spline curve, \eqref{eq:curve}\\
$\Nper$ & periodic collocation matrix, \eqref{eq:matrix}\\
$\mathbf K$ & stiffness matrix of the fairness energy, \eqref{eq:stiffness}\\
$\alpha,\beta$ & stretching and bending weights of $F$\\
$\theta_i$, $\boldsymbol\Theta$, $\mathbf x$ & knot coefficients, their vector, and unconstrained variables with $\theta_i=\sigma(x_i)$\\
$F(\boldsymbol\Theta)$ & fairness energy of a knot vector, \eqref{eq:fairness}\\
Opt-R, Opt-L & optimized right/left interlacing (Algorithm~\ref{alg:fairness})\\
shifting method & Park's shifting construction \cite{park2001choosing}\\
energy method & Park's constrained energy minimizer \cite{park2003lofted}\\
\bottomrule
\end{tabular}
\end{table}

Let $p$ be an even integer with $2\le p\le n$, and let $\mathbf Q_0,\ldots,\mathbf Q_n$ be
data points in a Euclidean space, with prescribed parameters
$\mathbf T(t_i)=\{t_0,\ldots,t_n\}$ satisfying
\begin{equation}
  0=t_0<t_1<\cdots<t_n<1.
  \label{eq:parameters}
\end{equation}
We seek a periodic B-spline curve interpolating these data at the
given parameters. The period is fixed at $1$. We set $t_{n+1}=1$
to describe the final parameter interval; it is not an additional
interpolation parameter.

Let the simple domain knots satisfy
\begin{equation}
  0=u_0<u_1<\cdots<u_n<u_{n+1}=1,
  \qquad u_{i+n+1}=u_i+1\quad(i\in\Z).
  \label{eq:knots}
\end{equation}
Let $N_{i,p}$ denote the normalized B-spline of degree $p$ associated
with the extended knot sequence \cite[Definition~4.19]{schumaker2007}.
We index these functions so that $N_{i,p}$ has support $[u_{i-p},u_{i+1}]$.
Define the periodic basis functions by
\begin{equation}
  \mathcal N_{j,p}(u)
  =\sum_{\ell\in\Z}N_{j+\ell(n+1),p}(u),
  \qquad 0\le j\le n.
  \label{eq:periodic-basis}
\end{equation}
The sum is locally finite. The periodic B-spline curve is
\begin{equation}
  \Ctilde(u)=\sum_{j=0}^n\mathcal N_{j,p}(u)\mathbf P_j,
  \qquad 0\le u\le1,
  \label{eq:curve}
\end{equation}
where $\mathbf P_0,\ldots,\mathbf P_n$ are the unknown control points.
The knot extension makes these functions $1$-periodic. For simple
knots, the ordinary B-splines are $C^{p-1}$ and have polynomial
pieces of degree at most $p$
\cite[Theorem~4.14]{schumaker2007}; the locally finite
sums in \eqref{eq:periodic-basis} inherit these properties.
In particular, the curve and its derivatives through order $p-1$
agree at the endpoints of the period.

The interpolation conditions are
\begin{equation}
  \Ctilde(t_i)=\mathbf Q_i,\qquad 0\le i\le n.
  \label{eq:interpolation}
\end{equation}
Writing $\mathbf P=(\mathbf P_0,\ldots,\mathbf P_n)^{\mathsf T}$ and
$\mathbf Q=(\mathbf Q_0,\ldots,\mathbf Q_n)^{\mathsf T}$, these
conditions give the square system
\begin{equation}
  \Nper\mathbf P=\mathbf Q,\qquad
  \Nper=\bigl(\mathcal N_{j,p}(t_i)\bigr)_{0\le i,j\le n}.
  \label{eq:matrix}
\end{equation}
Unique control points exist for arbitrary data precisely when
$\Nper$ is nonsingular.

We will also use the linear independence of the periodic functions.
If $\sum_{j=0}^n a_j\mathcal N_{j,p}=0$, extending the coefficients
by $a_{i+n+1}=a_i$ gives the locally finite identity
\begin{equation}
  \sum_{j=0}^n a_j\mathcal N_{j,p}(u)
  =\sum_{i\in\Z}a_iN_{i,p}(u)=0.
  \label{eq:periodic-independence}
\end{equation}
On $(u_k,u_{k+1})$, the active ordinary B-splines are
$N_{k,p},\ldots,N_{k+p,p}$. Their local linear independence
\cite[Theorem~4.18]{schumaker2007}
forces $a_k=\cdots=a_{k+p}=0$. Varying $k$ proves that all
coefficients vanish.

\section{Interlacing domain knots and unique solvability}
\label{sec:interlacing}

We give a constructive sufficient condition on the domain knots that makes
the periodic collocation matrix nonsingular for every even degree. The
condition is an interlacing of the knots with the parameters: around the
period the two families alternate. We first treat the right ordering, in
which each interior knot follows its corresponding parameter, and then
obtain the left ordering by reflection.

\begin{theorem}[Right interlacing]\label{thm:main}
Let the parameters and domain knots satisfy
\eqref{eq:parameters} and \eqref{eq:knots}, and let $p$ be even with $2\le p\le n$.
If
\begin{equation}
  u_0=t_0=0,\qquad
  t_i<u_i<t_{i+1}\quad(1\le i\le n),\qquad u_{n+1}=1,
  \label{eq:interlacing}
\end{equation}
then $\Nper$ is nonsingular. Consequently, for arbitrary data
$\mathbf Q_0,\ldots,\mathbf Q_n$, there is exactly one curve of the
form \eqref{eq:curve} satisfying \eqref{eq:interpolation}.
\end{theorem}

\begin{figure}[htbp]
  \centering
  \includegraphics[width=0.82\linewidth]{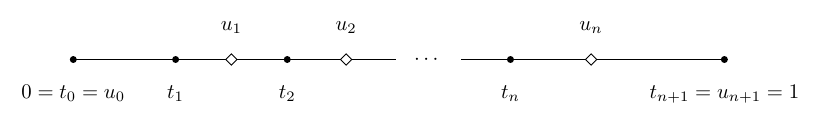}
  \caption{Interlacing with a common initial knot and parameter.
  Filled circles are interpolation parameters, and open diamonds
  are interior domain knots. The endpoints are identified by
  periodicity.}
  \label{fig:interlacing}
\end{figure}

Figure~\ref{fig:interlacing} illustrates the ordering in
\eqref{eq:interlacing}.

The proof, which uses two auxiliary lemmas, is given in
Appendix~\ref{app:proofs}.

The knot condition can be met directly by choosing any
$0<\theta_i<1$ and setting
\begin{equation}
  u_i=t_i+\theta_i(t_{i+1}-t_i),\qquad 1\le i\le n,
  \label{eq:knot-choice}
\end{equation}
with $u_0=0$ and $u_{n+1}=1$. Every such choice satisfies
Theorem~\ref{thm:main}.

The same conclusion holds for the opposite ordering, in which every
interior knot precedes its corresponding parameter.

\begin{corollary}[Left interlacing]\label{cor:left}
The conclusion of Theorem~\ref{thm:main} also holds when
\eqref{eq:interlacing} is replaced by
\begin{equation}
  u_0=t_0=0,\qquad
  t_{i-1}<u_i<t_i\quad(1\le i\le n),\qquad u_{n+1}=1.
  \label{eq:left-interlacing}
\end{equation}
\end{corollary}

\begin{figure}[H]
  \centering
  \includegraphics[width=0.82\linewidth]{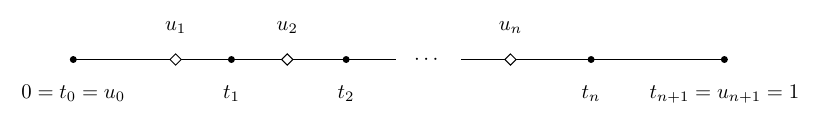}
  \caption{Left interlacing: each interior domain knot $u_i$ lies
  strictly between $t_{i-1}$ and $t_i$. Filled circles are interpolation
  parameters, and open diamonds are interior domain knots. The endpoints
  are identified by periodicity.}
  \label{fig:left-interlacing}
\end{figure}

The same left interlacing condition \eqref{eq:left-interlacing} was previously
put forward, as Conjecture~1, by Tang \cite{tang2022optimal} in the context of
lofted B-spline surface interpolation to serial closed contours, where it was
supported by numerical experiments but not proved. Corollary~\ref{cor:left}
establishes it rigorously for every even degree $p\ge2$.
The left ordering \eqref{eq:left-interlacing} is depicted in
Figure~\ref{fig:left-interlacing}.

\begin{proof}
Reflect the parameter by $u\mapsto1-u$. Set
$\widehat t_0=\widehat u_0=0$,
$\widehat t_{n+1}=\widehat u_{n+1}=1$, and
\begin{equation*}
  \widehat t_i=1-t_{n+1-i},\qquad
  \widehat u_i=1-u_{n+1-i},\qquad 1\le i\le n.
  \label{eq:reflection}
\end{equation*}
Then \eqref{eq:left-interlacing} implies
$\widehat t_i<\widehat u_i<\widehat t_{i+1}$ for $1\le i\le n$.

Take $\mathbf a$ with $\Nper\mathbf a=0$, and let
$s=\sum_{j=0}^n a_j\mathcal N_{j,p}$. The function
$\widehat s(v)=s(1-v)$ is periodic and $C^{p-1}$, has polynomial
pieces of degree at most $p$ on the reflected knot intervals,
and vanishes at every $\widehat t_i$. The homogeneous-spline
argument in the proof of Theorem~\ref{thm:main} therefore gives
$\widehat s\equiv0$, hence $s\equiv0$. Linear independence of the
original periodic basis gives $\mathbf a=0$. Thus $\Nper$ is
nonsingular, and the interpolation conclusion follows.
\end{proof}

In this ordering the direct knot choices are
\begin{equation}
  u_i=t_{i-1}+\theta_i(t_i-t_{i-1}),\qquad
  0<\theta_i<1,\qquad 1\le i\le n,
  \label{eq:left-knot-choice}
\end{equation}
with the same endpoints. Both results establish unique
solvability for each admissible choice.

\section{Fairness-optimized knot placement}
\label{sec:fairness}

The interlacing condition guarantees unique solvability for a whole family of
knot vectors, but the resulting curves differ in quality. A standard way to
measure this quality is the fairness (strain) energy of the curve~\cite{brunnett1994interpolation,vassilev1996fair}, 
a quadratic functional commonly used to penalize stretching and bending and
to favor smoother, better distributed curves. We select admissible
interlacing knots by locally optimizing this energy, which yields two
fairness-optimized constructions: Opt-R for the right ordering and Opt-L for
the left ordering. Each interior knot is determined by a single
convex-combination weight $\theta_i$, $0<\theta_i<1$
($1\le i\le n$):
\begin{align}
  \text{Opt-R:}\qquad & u_i=t_i+\theta_i\,(t_{i+1}-t_i),\qquad 1\le i\le n,\label{eq:theta-right}\\
  \text{Opt-L:}\qquad  & u_i=t_{i-1}+\theta_i\,(t_i-t_{i-1}),\qquad 1\le i\le n,\label{eq:theta-left}
\end{align}
with $u_0=t_0=0$ and $u_{n+1}=t_{n+1}=1$. For a given knot vector the curve is uniquely
determined by \eqref{eq:matrix}, and the fairness energy can be written
in terms of the control points and a stiffness matrix. Let
$\mathbf b(u)=(\mathcal N_{0,p}(u),\ldots,\mathcal N_{n,p}(u))^{\mathsf T}$
be the periodic basis vector, with first and second derivatives
$\mathbf b^{(1)}$ and $\mathbf b^{(2)}$. Then
\begin{equation}
  F(\boldsymbol\Theta)=\alpha\int_0^1\lVert\Ctilde'(u)\rVert^2\,du
             +\beta\int_0^1\lVert\Ctilde''(u)\rVert^2\,du
  =\operatorname{tr}\bigl(\mathbf P^{\mathsf T}\mathbf K\mathbf P\bigr),
  \label{eq:fairness}
\end{equation}
where
\begin{equation}
  \mathbf K=\int_0^1\bigl(\alpha\,\mathbf b^{(1)}(u)\mathbf b^{(1)}(u)^{\mathsf T}
        +\beta\,\mathbf b^{(2)}(u)\mathbf b^{(2)}(u)^{\mathsf T}\bigr)\,du
  \label{eq:stiffness}
\end{equation}
is the stiffness matrix, computed by Gaussian quadrature or analytically
\cite{moore2014efficient}; $\mathbf P=\Nper^{-1}\mathbf Q$ by
\eqref{eq:matrix}, and $\alpha\ge0$ and $\beta\ge0$ are the stretching and
bending coefficients \cite{park2001choosing,park2003lofted}; we use
$\alpha=1$ and $\beta=0.2$. 

Both $\mathbf K$ and $\Nper$ depend on the knot
vector, hence on $\boldsymbol\Theta$, so $F(\boldsymbol\Theta)$ is a function
of $\boldsymbol\Theta=(\theta_1,\ldots,\theta_n)$ only.
Although the fairness energy is quadratic in the control points for fixed
knots, the matrices $\Nper(\boldsymbol\Theta)$ and
$\mathbf K(\boldsymbol\Theta)$ depend nonlinearly on the knot coefficients.
The resulting finite-dimensional problem is therefore generally nonconvex,
so the numerical procedure below is a local optimization strategy. We search
for an admissible knot vector with $0<\theta_i<1$ ($1\le i\le n$). The box
constraint is removed by the logistic substitution
$\theta_i=1/(1+e^{-x_i})$ with $x_i\in\R$, and the transformed objective is
treated with the quasi-Newton (BFGS) method~\cite{nocedal2006numerical},
initialized from the two empirical values $\theta_i=0.5$ and $\theta_i=0.3$.
Among the candidates
returned by these runs, the one with the smallest computed energy is
retained. Because
$\theta_i\in(0,1)$ is preserved at every iterate, the interlacing
conditions \eqref{eq:theta-right}--\eqref{eq:theta-left} hold throughout, so the
interpolation matrix is always nonsingular. The procedure is summarized in
Algorithm~\ref{alg:fairness}, where $\delta$ selects the ordering.

\vspace{1.5mm}
\begin{breakablealgorithm}{Fairness-optimized interlacing knots
($\delta=\mathrm{right}$: Opt-R; $\delta=\mathrm{left}$: Opt-L)\label{alg:fairness}}
\begin{algorithmic}[1]
\Require data points $\mathbf Q_0,\ldots,\mathbf Q_n$; even degree $p$,
$2\le p\le n$; weights $\alpha,\beta$; direction
$\delta\in\{\mathrm{right},\mathrm{left}\}$
\Ensure admissible knots $u_0<\cdots<u_{n+1}$, control points
$\mathbf P$, interpolating curve $\Ctilde$, best computed energy
$F(\boldsymbol\Theta^\star)$
\State compute parameters $0=t_0<\cdots<t_n<1$; set $t_{n+1}=1$
\State set $\sigma(s)=1/(1+e^{-s})$, $u_0\gets0$, $u_{n+1}\gets1$
\State for $1\le i\le n$, set
$u_i(\boldsymbol\Theta)\gets
\begin{cases}
  t_i+\theta_i\,(t_{i+1}-t_i), & \delta=\mathrm{right},\\[2pt]
  t_{i-1}+\theta_i\,(t_i-t_{i-1}), & \delta=\mathrm{left}
\end{cases}$
\Function{$F$}{$\mathbf x$}\Comment{fairness energy of the interpolant}
  \State $\boldsymbol\Theta\gets\sigma(\mathbf x)$
  \State build $\mathbf u(\boldsymbol\Theta)$, $\Nper(\boldsymbol\Theta)$,
  $\mathbf K(\boldsymbol\Theta)$
  \Comment{$\mathbf u=(u_0,\ldots,u_{n+1})$: domain knots}
  \State solve $\Nper\mathbf P=\mathbf Q$
  \State \Return $\operatorname{tr}(\mathbf P^{\mathsf T}\mathbf K\mathbf P)$
\EndFunction
\State $\mathcal I\gets\{0.5,\ 0.3\}$ \Comment{empirical initial values of $\theta_i$}
\For{$\theta^{(0)}\in\mathcal I$}
  \State $\boldsymbol\Theta^{(0)}\gets(\theta^{(0)},\ldots,\theta^{(0)})$
  \State $\mathbf x^\star\gets
  \operatorname{BFGS}\bigl(F,\operatorname{logit}(\boldsymbol\Theta^{(0)})\bigr)$
  \Comment{$\operatorname{logit}=\sigma^{-1}$}
  \State $\boldsymbol\Theta^\star\gets\sigma(\mathbf x^\star)$
\EndFor
\State retain, among the computed candidates, the $\boldsymbol\Theta^\star$
with the smallest computed value of $F$
\State build $\mathbf u^\star\gets\mathbf u(\boldsymbol\Theta^\star)$;
solve $\Nper(\mathbf u^\star)\mathbf P^\star=\mathbf Q$
\State \Return $\mathbf u^\star,\ \mathbf P^\star,\
\Ctilde(u)=\mathbf b(u)^{\mathsf T}\mathbf P^\star,\
F(\boldsymbol\Theta^\star)$
\end{algorithmic}
\end{breakablealgorithm}

\section{Experimental results}
\label{sec:experiments}

\begin{figure}[htbp]
  \centering
  \includegraphics[width=\textwidth]{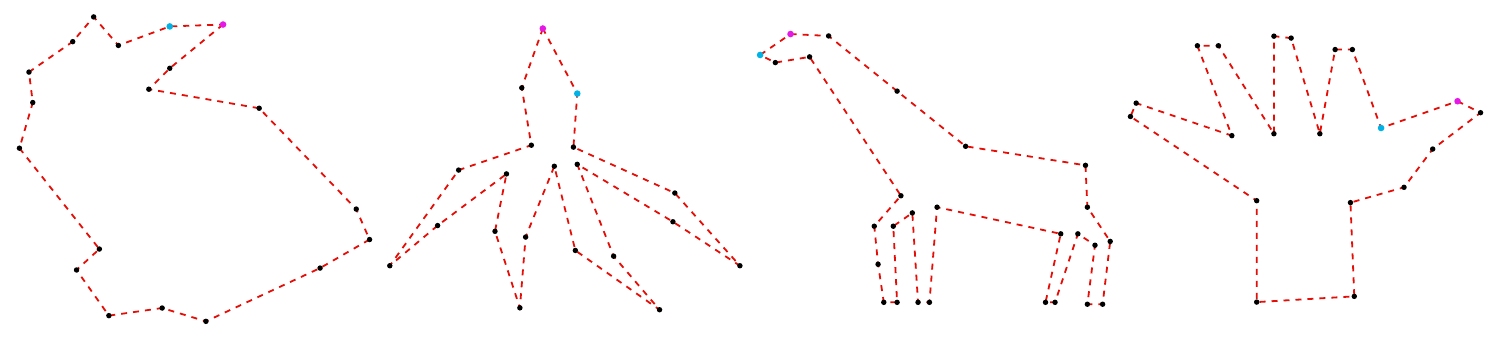}
  \caption{The four data sets used in the experiments: rabbit ($19$ vertices,
  degree $p=2$), octopus ($20$ vertices, $p=4$), giraffe ($27$ vertices,
  $p=2$), and fingers ($20$ vertices, $p=4$). Each data set is normalized to
  unit bounding-box diagonal; the first and last vertices are marked in
  magenta and cyan, respectively.}
  \label{fig:data-sets}
\end{figure}

The methods are implemented in C and Wolfram Language. This section evaluates
the optimized interlacing methods on four closed polygonal data sets with
complex, nonuniform point distributions, at degrees $p=2$ and $p=4$. 
The data sets, referred to as rabbit, octopus, giraffe, and
fingers and shown in Figure~\ref{fig:data-sets}, are each centered at the
origin and scaled to unit bounding-box diagonal, so that the curvature-based
quantities are comparable across shapes. The first two, rabbit and octopus,
have $19$ and $20$ vertices and are processed at degrees $p=2$ and $p=4$,
respectively. The third, giraffe,
has $27$ vertices and is processed at $p=2$; with several sharp corners, it is
the most irregular of the four. The fourth, fingers, has $20$ vertices and is
processed at $p=4$.

Every data set is interpolated with three parametrizations, namely the
classical chord-length and centripetal methods \cite{lee1989choosing} and the
improved parametrization of Fang and Hung \cite{fang2013improved} (denoted
fang2013), and with four constructions: the optimized interlacing methods
Opt-R and Opt-L of Section~\ref{sec:fairness}, together with Park's shifting
and energy methods \cite{park2001choosing}. Appendix~\ref{app:compared}
recalls the two deterministic constructions, Section~\ref{subsec:quality}
defines the quality metric, and the rest of this section reports the results.

\subsection{Quality metric}
\label{subsec:quality}

To evaluate the quality of an interpolation curve we use the metric of Fang
and Hung \cite{fang2013improved},
\begin{equation}
  W_C=\frac{1}{n+1}\sum_{i=0}^{n}(1+K_i)\,D_i,
  \label{eq:wc}
\end{equation}
where $K_i$ is the maximum absolute curvature of the curve on the $i$th
parameter interval and $D_i$ is the maximum deviation of the curve from the
corresponding data segment,
\begin{equation}
  D_i=\max_{t_i<t<t_{i+1}}\operatorname{dist}\bigl(\Ctilde(t),L_i\bigr),
  \qquad L_i=\overline{\mathbf Q_i\mathbf Q_{i+1}},
  \label{eq:wd}
\end{equation}
with $\operatorname{dist}(\Ctilde(t),L_i)$ the distance from the curve point
$\Ctilde(t)$ to the segment $L_i$. For a periodic curve the data are closed and
there are $n+1$ segments: the last interval $(t_n,t_{n+1})$, with $t_{n+1}=1$,
corresponds to the closing segment $L_n=\overline{\mathbf Q_n\mathbf Q_0}$, and
the index in \eqref{eq:wd} is taken modulo $n+1$. A smaller $W_C$ indicates a
curve that follows the data polyline closely while bending little.

\subsection{Reconstructed curves and knot placement}
\label{subsec:curves}

\begin{figure}[htbp]
  \centering
  \includegraphics[width=0.90\textwidth]{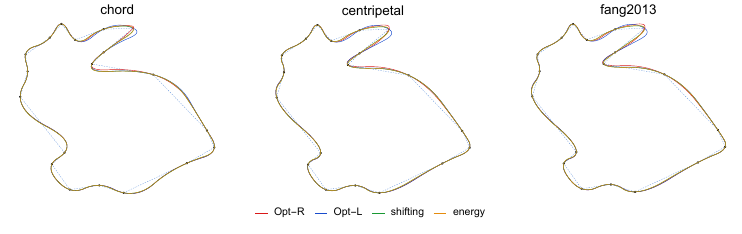}\\[1mm]
  \makebox[\textwidth][c]{\includegraphics[width=1.05\textwidth]{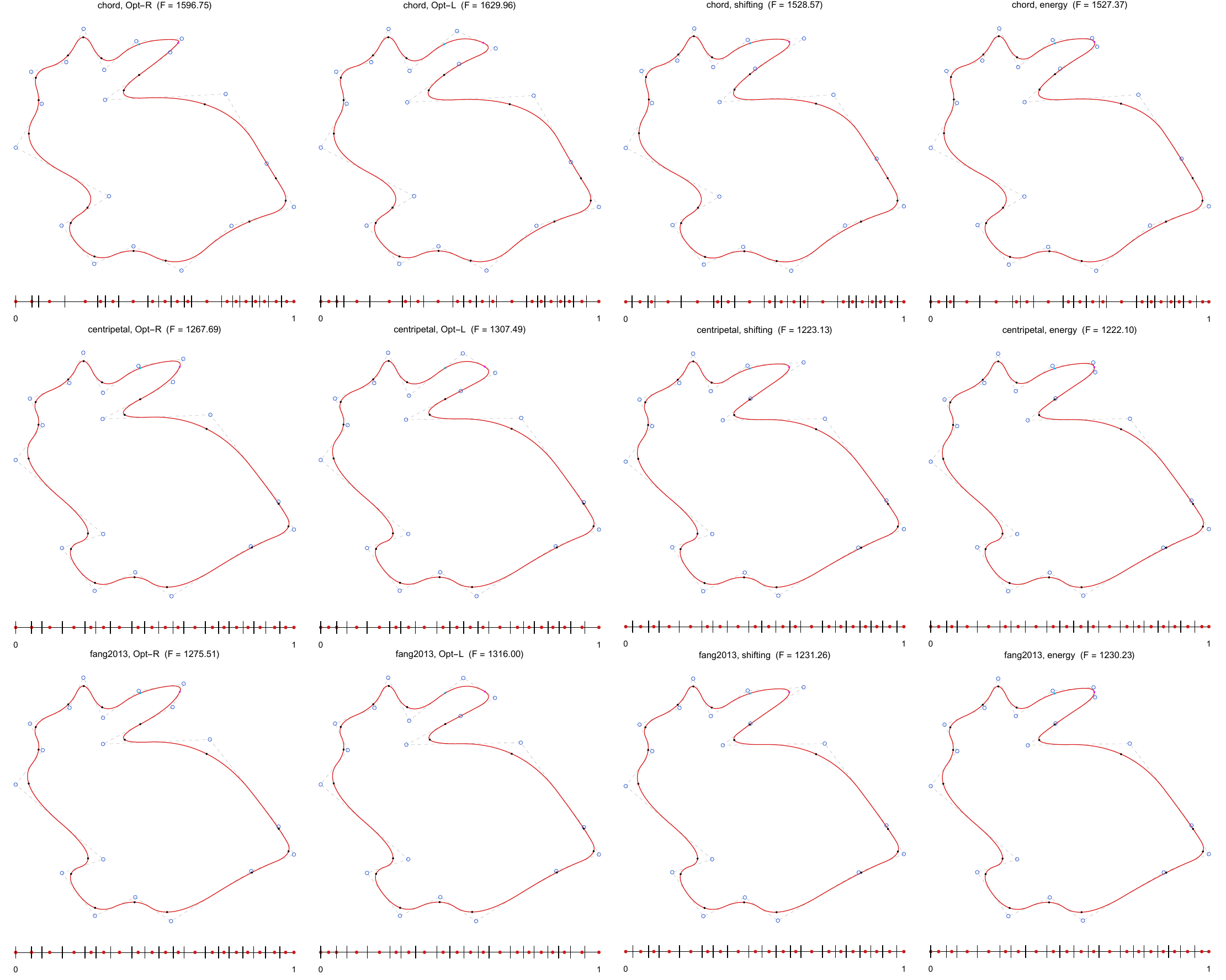}}\\[1mm]
  \includegraphics[width=0.75\textwidth]{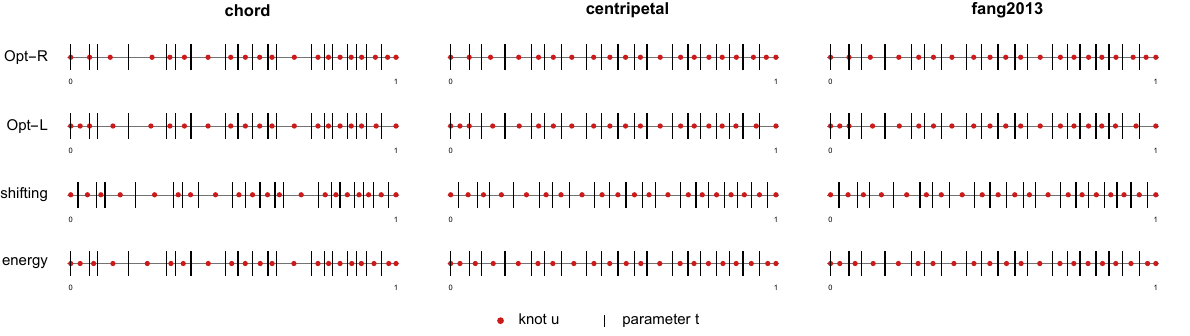}
  \caption{rabbit ($p=2$, $19$ vertices): interpolants of the four
  constructions under the three parametrizations (top), reconstructions with
  control polygons and data points (middle), and knot/parameter placement
  (bottom).}
  \label{fig:data-1}
\end{figure}

\begin{figure}[htbp]
  \centering
  \includegraphics[width=0.90\textwidth]{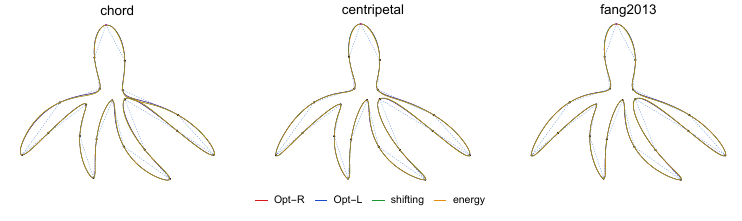}\\[1mm]
  \makebox[\textwidth][c]{\includegraphics[width=1.05\textwidth]{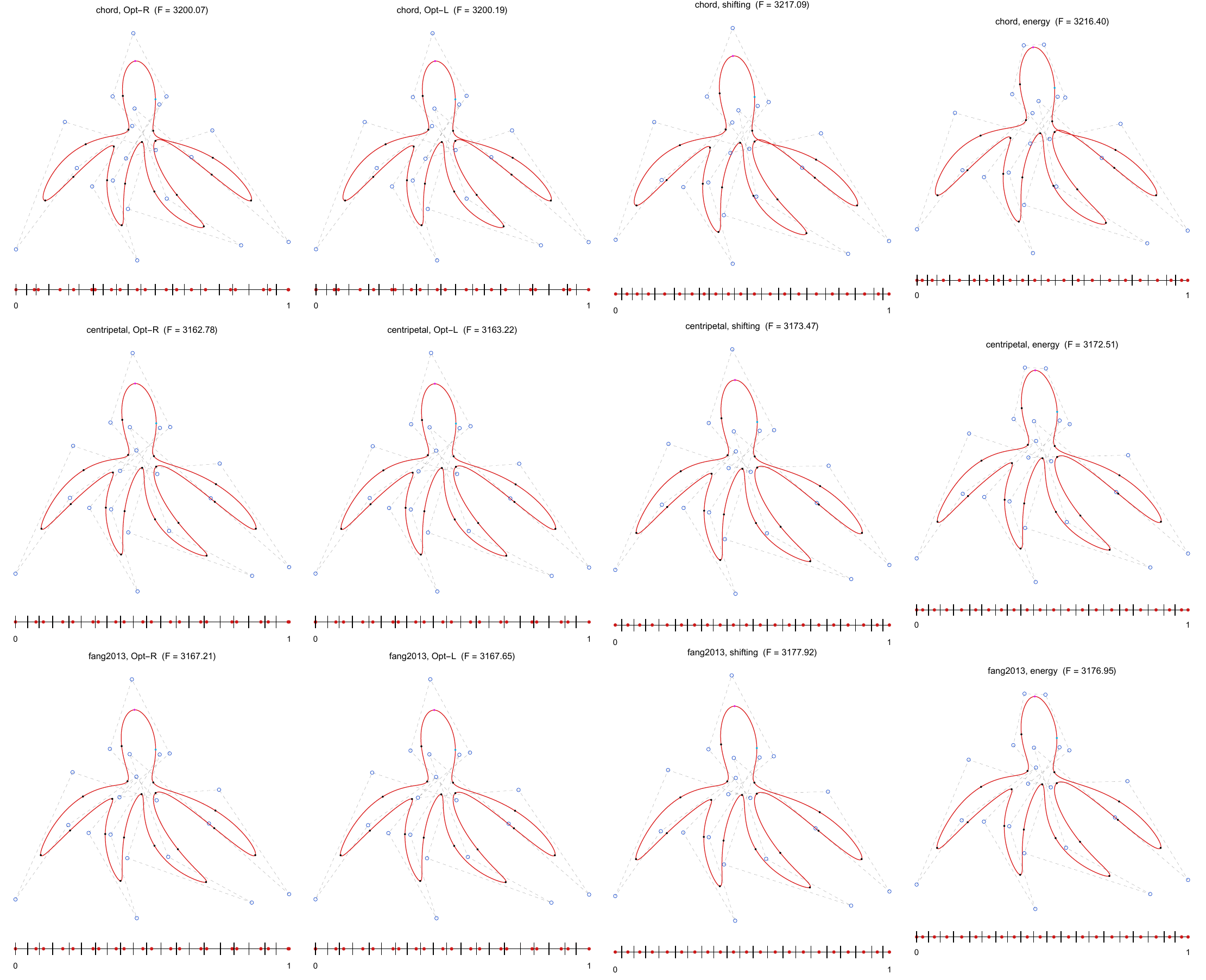}}\\[1mm]
  \includegraphics[width=0.85\textwidth]{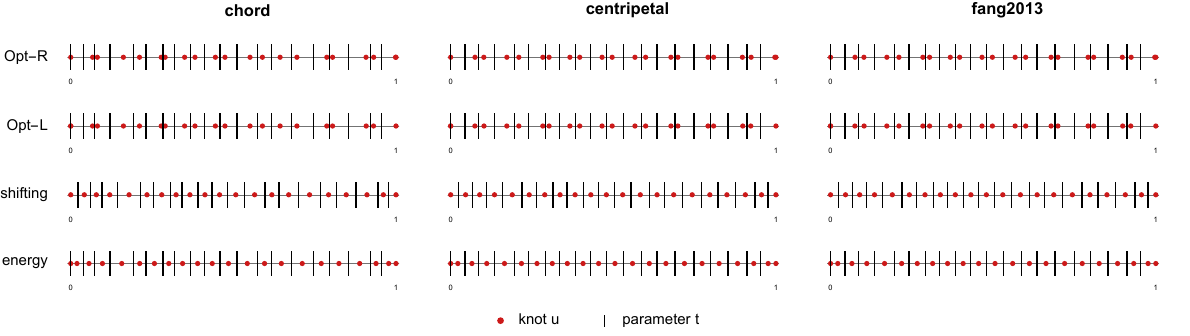}
  \caption{octopus ($p=4$, $20$ vertices): interpolants of the four
  constructions under the three parametrizations (top), reconstructions with
  control polygons and data points (middle), and knot/parameter placement
  (bottom).}
  \label{fig:data-2}
\end{figure}

\begin{figure}[htbp]
  \centering
  \includegraphics[width=0.94\textwidth]{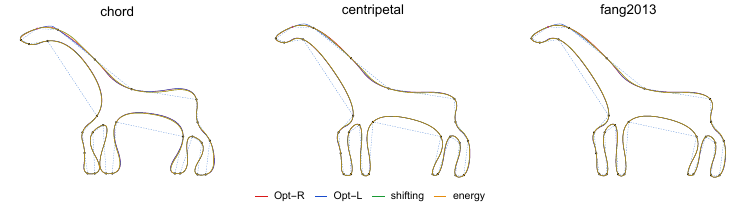}\\[1mm]
  \makebox[\textwidth][c]{\includegraphics[width=1.05\textwidth]{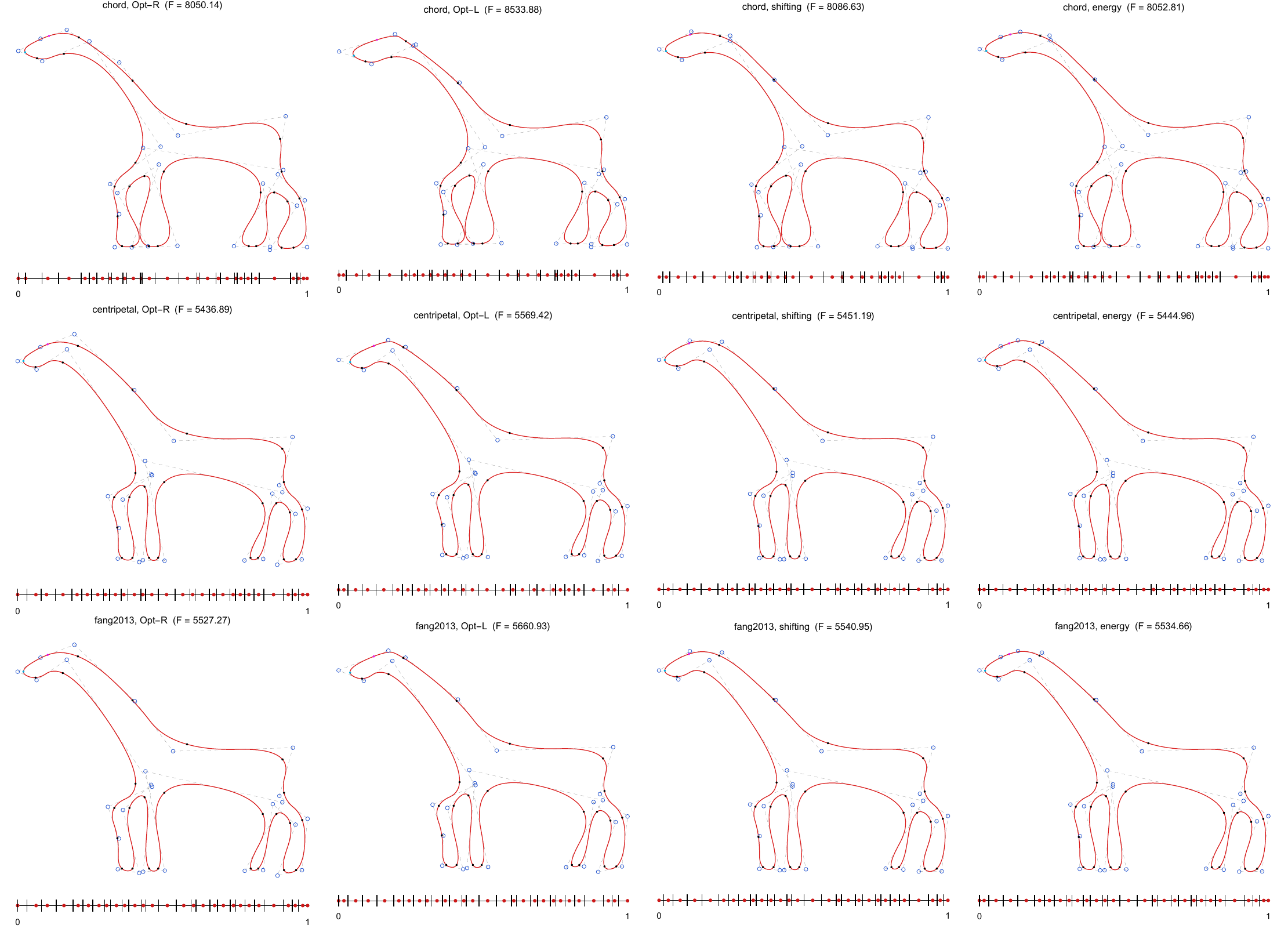}}\\[1mm]
  \includegraphics[width=0.75\textwidth]{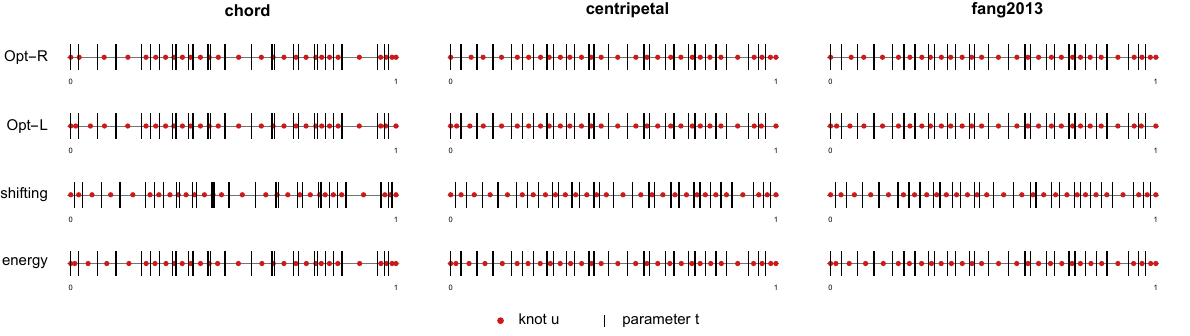}
  \caption{giraffe ($p=2$, $27$ vertices): interpolants of the four
  constructions under the three parametrizations (top), reconstructions with
  control polygons and data points (middle), and knot/parameter placement
  (bottom).}
  \label{fig:data-3}
\end{figure}

\begin{figure}[htbp]
  \centering
  \includegraphics[width=0.94\textwidth]{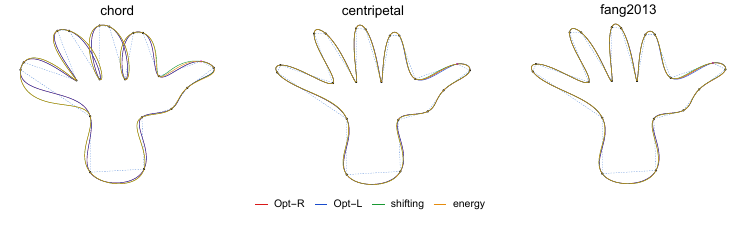}\\[1mm]
  \makebox[\textwidth][c]{\includegraphics[width=1.05\textwidth]{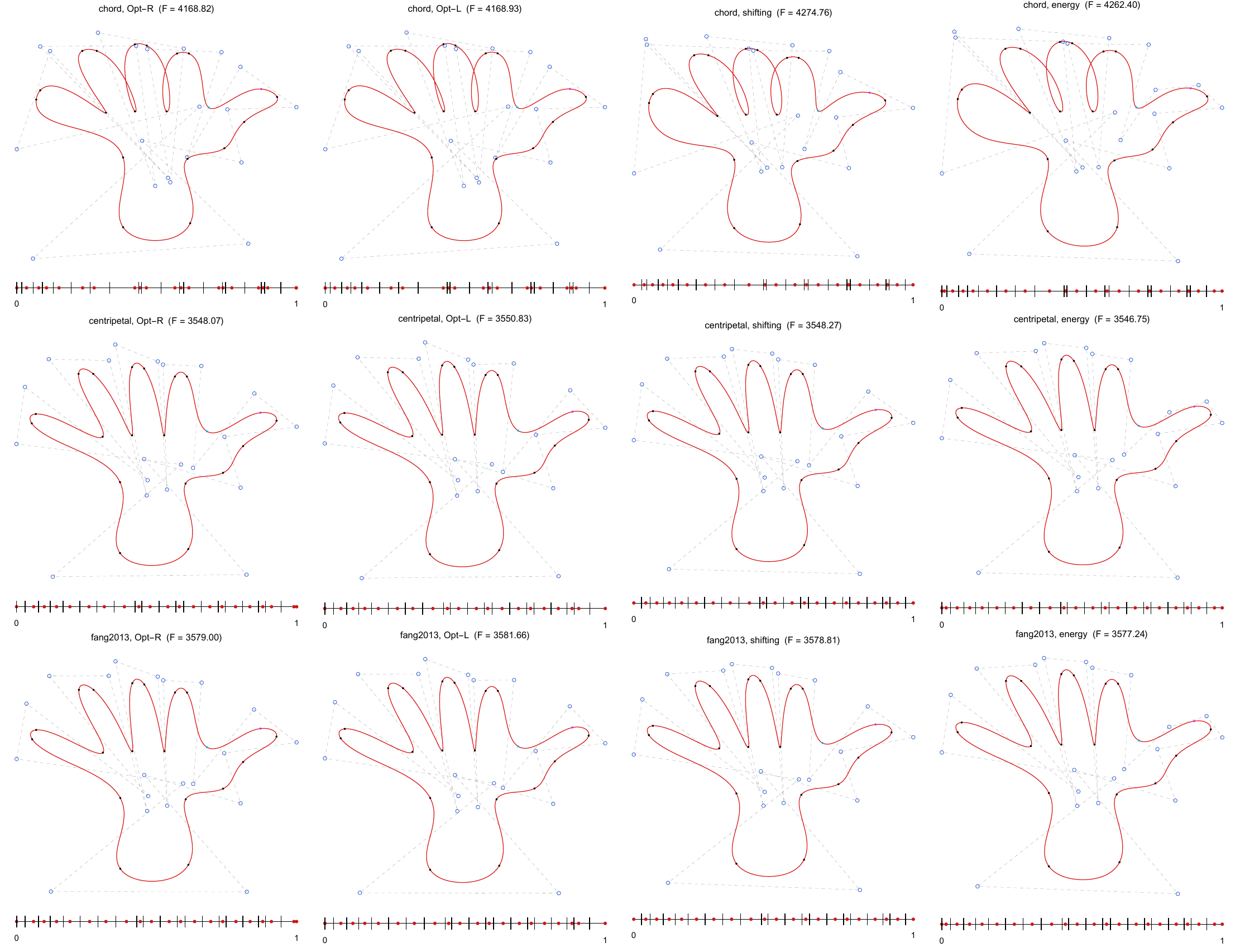}}\\[1mm]
  \includegraphics[width=0.75\textwidth]{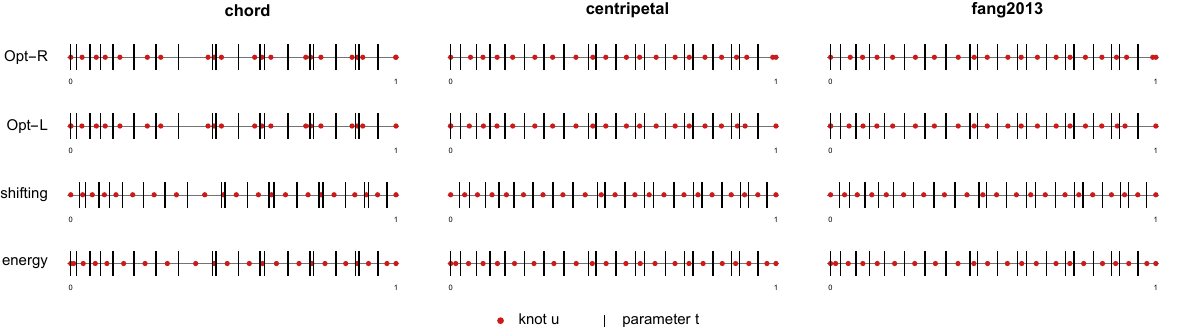}
  \caption{fingers ($p=4$, $20$ vertices): interpolants of the four
  constructions under the three parametrizations (top), reconstructions with
  control polygons and data points (middle), and knot/parameter placement
  (bottom).}
  \label{fig:data-4}
\end{figure}

\begin{figure}[H]
  \centering
  \includegraphics[width=0.9\textwidth]{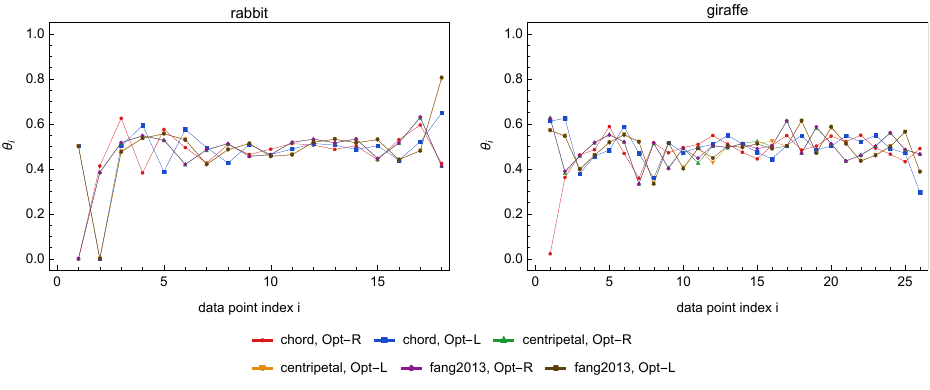}\\[1mm]
  \includegraphics[width=0.9\textwidth]{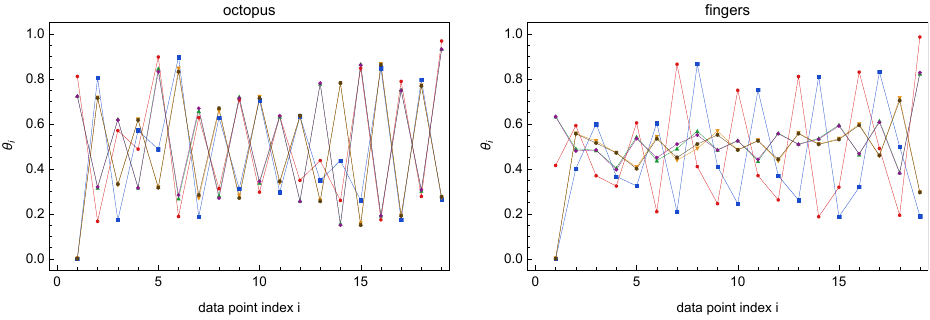}
  \caption{Optimized knot coefficients $\theta_i$ versus the data-point index
  for the two interlacing methods. Top, degree $p=2$: rabbit (left)
  and giraffe (right); bottom, degree $p=4$: octopus (left)
  and fingers (right).}
  \label{fig:theta}
\end{figure}

In each of Figures~\ref{fig:data-1}--\ref{fig:data-4} the top panel
overlays the four interpolants obtained for each data set and parametrization.
All constructions interpolate the data exactly, and their curves are visually
indistinguishable along the smooth parts of the polygons; they differ only
near highly curved vertices, where the placement of the interior knots
determines how early the curve has to turn. This confirms that the freedom
left by the interlacing condition is an effective degree of freedom that the four
constructions resolve in different ways. The middle panel shows the full
reconstructions, including the control polygons and the data points, with the
fairness energy $F$ of each construction printed above it, and the bottom panel
displays the resulting knot and parameter placement: the \emph{shifting}
and \emph{energy} methods place their knots at the midpoints of consecutive
parameters, while Opt-R and Opt-L distribute them nonuniformly according to
$\theta$ (Figure~\ref{fig:theta}).

Two of the data sets deserve particular attention. In
Figure~\ref{fig:data-2} (octopus) the optimized interlacing is clearly
superior: Opt-R and Opt-L produce visually pleasing curves, whereas the
centered shifting and energy constructions exhibit small self-intersections
under chord-length parametrization. Since all four constructions use the same
parameters and differ only in the
placement of the interior knots, this comparison isolates the effect of knot
placement and shows that the optimized interlacing avoids the artifacts
introduced by centered knots. Because the metric \eqref{eq:wc} does not
penalize self-intersections, this visual improvement is not reflected in
Table~\ref{tab:wc-paper}, where the centered knots can show a smaller
$W_C$. In Figure~\ref{fig:data-4} (fingers), by
contrast, all four constructions produce self-intersecting curves under
chord-length parametrization. Here the interior knots are no
longer the deciding factor: the artifact originates in the parametrization,
and no admissible knot placement remedies it. This is the same phenomenon
that occurs in open-curve interpolation, where the parametrization governs the
shape and, in particular, the appearance of self-intersections. The two cases
therefore complement each other and illustrate the interplay between
parametrization and knot placement: knot optimization is effective when the
knots are the limiting factor, whereas a poor parametrization cannot be
compensated by any knot placement.

Figure~\ref{fig:theta} shows the optimized knot coefficients $\theta_i$ of
Opt-R and Opt-L. Their values vary substantially from vertex to vertex and
track the local geometry: for the irregular data set giraffe the
coefficients range over nearly the whole admissible interval $[0,1]$, whereas
for the smoother sets they remain closer to the midpoint $1/2$. The right and
left orderings produce visibly different patterns, reflecting that they move
each knot against the two different bounding parameters. For the centripetal
and fang2013 parametrizations the two $\theta_i$ curves in
Figure~\ref{fig:theta} are very close to each other but not identical. This is a geometric effect rather
than a numerical coincidence: the two parametrizations themselves differ only
slightly, and near the optimum a small change of a knot has only a second-order
effect on the curve, so both parametrizations drive the coefficients to almost
the same values. The resulting curves agree to about $0.1\%$ of the shape size,
even though the underlying parameter vectors differ by up to
$7.2\times10^{-3}$ and the fairness energy still differs by roughly $1.5\%$.

\begin{table}[htbp]
\centering\small
\caption{Condition numbers $\kappa_2(\Nper)$ of the interpolation systems for the four data sets.}
\label{tab:cond-paper}
\begin{tabular}{llcccc}
\toprule
Data & Param. & Opt-R & Opt-L & shifting & energy \\
\midrule
  \multirow{3}{*}{rabbit ($p=2$)} & chord & 4.047 & 3.032 & 3.085 & 3.083 \\
   & centripetal & 3.363 & 2.670 & 2.178 & 2.166 \\
   & fang2013 & 3.359 & 2.683 & 2.177 & 2.164 \\
  \addlinespace
  \multirow{3}{*}{octopus ($p=4$)} & chord & 5.494 & 5.501 & 5.353 & 5.029 \\
   & centripetal & 5.104 & 5.121 & 4.913 & 4.756 \\
   & fang2013 & 5.100 & 5.119 & 4.911 & 4.756 \\
  \addlinespace
  \multirow{3}{*}{giraffe ($p=2$)} & chord & 5.711 & 6.141 & 5.620 & 5.554 \\
   & centripetal & 2.566 & 2.921 & 2.575 & 2.554 \\
   & fang2013 & 2.581 & 2.928 & 2.587 & 2.567 \\
  \addlinespace
  \multirow{3}{*}{fingers ($p=4$)} & chord & 11.976 & 11.980 & 14.621 & 14.536 \\
   & centripetal & 6.410 & 6.425 & 6.292 & 6.260 \\
   & fang2013 & 6.445 & 6.460 & 6.381 & 6.352 \\
\bottomrule
\end{tabular}
\end{table}

Table~\ref{tab:cond-paper} lists the condition numbers $\kappa_2(\Nper)$ of
the periodic collocation matrices. The parametrization has the strongest
influence: chord-length yields the largest condition numbers, whereas
centripetal and fang2013 reduce them by roughly a factor of two, for example
from $5.71$ to $2.57$ for giraffe and from $11.98$ to $6.41$ for
fingers. The four constructions have condition numbers of the same
order; the optimized interlacing methods Opt-R and Opt-L are close to the
deterministic shifting and energy constructions. All values remain moderate,
the largest being about $14.6$ for fingers under chord-length, which is
consistent with the nonsingularity guaranteed by Theorem~\ref{thm:main} and
indicates that the admissible knot placements are numerically safe.


The fairness energy $F$ is collected in Table~\ref{tab:energy-paper}. All four
constructions give values within a few percent of one another, and the smallest
value alternates between the energy method and the optimized methods.
Differences between the parametrizations are larger: relative to chord-length,
centripetal reduces $F$ by about a third for giraffe ($5437$ versus $8050$)
and by roughly a fifth for rabbit and a seventh for fingers, while for octopus the three
parametrizations give similar values.

Table~\ref{tab:wc-paper} shows the Fang--Hung metric $W_C$ of
\eqref{eq:wc}; smaller values indicate a curve that follows the data polygon
closely while bending little. The \emph{shifting} and \emph{energy} methods attain the
smallest values in most cases, which reflects that centered knots keep the
deviation from the chords small for well-distributed parameters; the optimized
methods are close to this baseline, within about fifteen percent for
octopus, within a few percent for giraffe, and within about a quarter for rabbit. The notable exception is
fingers under chord-length parametrization, where the centered
constructions (shifting or energy) give $W_C\approx10.7$, more than twice the value
$W_C\approx5.24$ of Opt-R and Opt-L. Here the strong nonuniformity of the
chord-length parameters makes the centered knots deviate appreciably from the
data polygon, whereas the optimized interlacing adapts the knots to the
geometry and keeps the deviation small. Centripetal and fang2013 eliminate
most of this discrepancy for fingers ($W_C\approx2.5$ to $3.7$). Thus
no method is uniformly best: the parametrization and the knot placement
interact, and the optimized interlacing provides a robust compromise.

\begin{table}[H]
\centering\small
\caption{Fairness energy $F$ (smaller is better) of the interpolation systems for the four data sets.}
\label{tab:energy-paper}
\begin{tabular}{llcccc}
\toprule
Data & Param. & Opt-R & Opt-L & shifting & energy \\
\midrule
  \multirow{3}{*}{rabbit ($p=2$)} & chord & 1596.75 & 1629.96 & 1528.57 & \textbf{1527.37} \\
   & centripetal & 1267.69 & 1307.49 & 1223.13 & \textbf{1222.10} \\
   & fang2013 & 1275.51 & 1316.00 & 1231.26 & \textbf{1230.23} \\
  \addlinespace
  \multirow{3}{*}{octopus ($p=4$)} & chord & \textbf{3200.07} & 3200.19 & 3217.09 & 3216.40 \\
   & centripetal & \textbf{3162.78} & 3163.22 & 3173.47 & 3172.51 \\
   & fang2013 & \textbf{3167.21} & 3167.65 & 3177.92 & 3176.95 \\
  \addlinespace
  \multirow{3}{*}{giraffe ($p=2$)} & chord & \textbf{8050.14} & 8533.88 & 8086.63 & 8052.81 \\
   & centripetal & \textbf{5436.89} & 5569.42 & 5451.19 & 5444.96 \\
   & fang2013 & \textbf{5527.27} & 5660.93 & 5540.95 & 5534.66 \\
  \addlinespace
  \multirow{3}{*}{fingers ($p=4$)} & chord & \textbf{4168.82} & 4168.93 & 4274.76 & 4262.40 \\
   & centripetal & 3548.07 & 3550.83 & 3548.27 & \textbf{3546.75} \\
   & fang2013 & 3579.00 & 3581.66 & 3578.81 & \textbf{3577.24} \\
\bottomrule
\end{tabular}
\end{table}

\begin{table}[htbp]
\centering\small
\caption{Fang--Hung quality metric $W_C$ (smaller is better) for the four data sets.}
\label{tab:wc-paper}
\begin{tabular}{llcccc}
\toprule
Data & Param. & Opt-R & Opt-L & shifting & energy \\
\midrule
  \multirow{3}{*}{rabbit ($p=2$)} & chord & 0.5560 & 0.5241 & \textbf{0.4371} & 0.4807 \\
   & centripetal & 0.5004 & 0.4710 & \textbf{0.4062} & 0.4612 \\
   & fang2013 & 0.5072 & 0.4706 & \textbf{0.4049} & 0.4630 \\
  \addlinespace
  \multirow{3}{*}{octopus ($p=4$)} & chord & 2.0718 & 2.0739 & \textbf{1.8873}$^{\dagger}$ & 1.9331$^{\dagger}$ \\
   & centripetal & 2.1914 & 2.1944 & \textbf{1.9398} & 1.9842 \\
   & fang2013 & 2.1706 & 2.1736 & \textbf{1.9344} & 1.9782 \\
  \addlinespace
  \multirow{3}{*}{giraffe ($p=2$)} & chord & 0.5485 & 0.5880 & \textbf{0.5350} & 0.5367 \\
   & centripetal & 0.4995 & 0.5237 & \textbf{0.4811} & 0.4820 \\
   & fang2013 & 0.4947 & 0.5206 & \textbf{0.4777} & 0.4788 \\
  \addlinespace
  \multirow{3}{*}{fingers ($p=4$)} & chord & \textbf{5.2360}$^{\dagger}$ & 5.2367$^{\dagger}$ & 10.7566$^{\dagger}$ & 10.6820$^{\dagger}$ \\
   & centripetal & 3.6007 & 3.6640 & \textbf{2.5748} & 2.6031 \\
   & fang2013 & 3.4824 & 3.5433 & \textbf{2.5384} & 2.5664 \\
\bottomrule
\end{tabular}

\par\vspace{2pt}
{\footnotesize $^{\dagger}$ marks an interpolation curve that self-intersects.}
\end{table}

\begin{table}[H]
\centering\small
\caption{Computation time (in milliseconds) per data set, parametrization and construction.}
\label{tab:time-paper}
\begin{tabular}{llcccc}
\toprule
Data & Param. & Opt-R & Opt-L & shifting & energy \\
\midrule
  \multirow{3}{*}{rabbit ($p=2$)} & chord & 75.18 & 78.69 & 0.32 & 0.73 \\
   & centripetal & 69.09 & 81.86 & 0.15 & 0.34 \\
   & fang2013 & 76.89 & 75.36 & 0.20 & 0.40 \\
  \addlinespace
  \multirow{3}{*}{octopus ($p=4$)} & chord & 68.73 & 69.40 & 0.22 & 0.50 \\
   & centripetal & 74.73 & 64.07 & 0.61 & 0.72 \\
   & fang2013 & 68.04 & 71.08 & 0.20 & 0.83 \\
  \addlinespace
  \multirow{3}{*}{giraffe ($p=2$)} & chord & 74.06 & 83.87 & 0.20 & 0.66 \\
   & centripetal & 86.74 & 68.62 & 0.24 & 0.54 \\
   & fang2013 & 67.21 & 72.08 & 0.25 & 0.55 \\
  \addlinespace
  \multirow{3}{*}{fingers ($p=4$)} & chord & 84.94 & 72.80 & 0.20 & 0.50 \\
   & centripetal & 76.06 & 73.84 & 0.40 & 0.92 \\
   & fang2013 & 97.70 & 93.34 & 0.27 & 0.53 \\
\bottomrule
\end{tabular}
\end{table}


Finally, Table~\ref{tab:time-paper} compares the computation time of each
construction.
The two deterministic constructions are the fastest: both reduce to
assembling and solving a single dense linear system and take well under a
millisecond, the energy method being marginally more expensive because it
solves a KKT system of about twice the size. The optimized methods Opt-R and
Opt-L are about two orders of magnitude slower, their time being dominated
by the multi-start optimization of the knot coefficients, but even for the
most expensive case they require only about $0.1$ s per parametrization. This
is already practical for real applications, and further code-level optimization
could reduce the running time even more.

Taken together, the experiments show that the interlacing constructions Opt-R
and Opt-L are uniquely solvable and well conditioned for all four data sets and
both degrees, with condition numbers comparable to those of the deterministic
shifting and energy methods and markedly smaller under centripetal and
fang2013 parametrization than under chord-length. Their curve quality,
measured by $W_C$, is close to the best deterministic construction on smooth
data and appreciably better on the difficult combination of fingers
with chord-length parametrization. The optimized knot placement therefore
recovers the geometric accuracy of the centered constructions (shifting or energy) while retaining
the flexibility to adapt to irregular data. Additional examples, covering
twelve further test shapes at degree $p=2$ with centripetal parametrization,
are given in Appendix~\ref{app:shapes}.

\section{Conclusions and future work}
\label{sec:conclusions}

We have shown that the interlacing of the domain knots with the
interpolation parameters is what makes the even-degree periodic
B-spline interpolation problem uniquely solvable. The condition is
constructive and mild: the endpoints are pinned, and every interior
knot only has to lie somewhere inside the interval between two
consecutive parameters. It therefore guarantees a nonsingular
collocation matrix for a whole family of knot vectors, independently
of how nonuniform the parameters are, and it covers both the right
and the left ordering of the knots.
The freedom left by the interlacing condition can be exploited
directly. Locally optimizing one coefficient per interior knot gives a
fairness-optimized knot placement procedure, implemented with a logistic
reparametrization and a multi-start quasi-Newton search. The
logistic substitution keeps every coefficient strictly inside the
open unit interval, so the interlacing condition, and with it unique
solvability, holds at every iterate of the optimization. The
comparison with the shifting and energy methods shows that the locally
optimized interlacing methods and the deterministic energy method achieve low
fairness energies, that the shifting method is a cheap and
well-conditioned alternative, and
that centripetal parametrization is preferable to chord-length
parametrization for the conditioning of the interpolation system. The shifting
construction, widely used without a proof of solvability, is also covered by the
interlacing theory: its collocation matrix is nonsingular, even though Park's
original diagonal-dominance argument does not hold
(Appendix~\ref{app:park-dominance}).

From a practical viewpoint the running time of the optimized methods is modest.
The deterministic shifting and energy constructions reduce to a single linear
solve and are about two orders of magnitude faster, but even in the most
expensive case Opt-R and Opt-L need only about $0.1$~s per parametrization. This
time is dominated by the multi-start search, so code-level optimization can
reduce it further, and a tenth of a second is already negligible for interactive
design and other typical CAGD tasks. The optimization therefore adds a guaranteed
solvability criterion and adaptivity to irregular data at a cost that remains well
within the practical budget.

Three directions stand out for future work. The first is to extend the
interlacing analysis to odd degree. The present theory covers even degrees,
for which the interlacing condition gives a constructive sufficient condition
for nonsingularity; for odd degrees the situation is open, since the periodic
window condition of Section~\ref{subsec:stability} fails in general
(Appendix~\ref{app:window-counterexample}). Identifying the window of
admissible positions for each domain knot around its parameter that guarantees
nonsingularity at odd degree would complete the solvability picture across all
degrees. The second direction is a forward construction of the coefficients.
Rather than selecting $\theta_i$ through a multi-start local search on the
fairness energy, it would be preferable to compute $\theta_i$ directly from
the data; a natural conjecture is that, at least for suitably regular data,
the geometric midpoint $\theta_i=1/2$ gives a reasonable or near-optimal
choice. Establishing such a result would reduce the construction to an
explicit formula. The third direction is the conditioning of the interpolation
system: although the experiments indicate that the admissible knot placements are
well conditioned, a rigorous bound on the condition number of the periodic
collocation matrix, and its dependence on the parametrization, remains to be
established. The same interlacing viewpoint
should also extend to periodic tensor-product B-spline surfaces and to
closed-contour lofting, where the point-wise method would be combined with a
fairing term across the contours.

\section*{Declaration of competing interest}
The authors declare that they have no known competing financial interests or personal relationships that could have appeared to influence the work reported in this paper.


\appendix

\section{Proofs for Section~\ref{sec:interlacing}}
\label{app:proofs}

\subsection{Auxiliary lemmas}

Distinct zeros
$x_1<\cdots<x_k<x_1+1$ of a continuous
$1$-periodic function $f$ are called \emph{separated} if $f$ is not
identically zero on any $(x_j,x_{j+1})$, where $x_{k+1}=x_1+1$.
This definition includes $k=1$; the identically zero function has
no nonempty set of separated zeros.

\begin{lemma}[Separated zeros]\label{lem:zeros}
Let $q\ge1$ be an integer. A $1$-periodic $C^{q-1}$ function whose restriction to
each knot interval $[u_i,u_{i+1}]$ is a polynomial of degree at most
$q$ has at most $n+1$ separated zeros in a period.
\end{lemma}

\begin{proof}[Proof of Lemma~\ref{lem:zeros}]
Suppose a periodic $C^1$ function $f$ has $k$ separated zeros $x_j$.
Between successive zeros choose an interior point $y_j$ maximizing
$|f|$. Then $f'(y_j)=0$ and $f(y_j)\ne0$.

With $y_{k+1}=y_1+1$, the point $x_{j+1}$ lies between
$y_j$ and $y_{j+1}$. Since $f(x_{j+1})=0$ and $f(y_j)\ne0$,
$f'$ is not identically zero
between successive $y_j$, so these are $k$ separated zeros of $f'$.

If a spline as in the statement has $k$ separated zeros, applying
this argument $q-1$ times gives $k$ separated zeros of its continuous
piecewise linear $(q-1)$st derivative. Between two consecutive such zeros there
must be a knot in the open interval: otherwise the function would
be affine there and zero at both endpoints, contrary to separation.
The $k$ open intervals are disjoint on the circle, which contains
only $n+1$ knots. Therefore $k\le n+1$.
\end{proof}

\begin{lemma}[An identity for even-degree periodic splines]
\label{lem:jump-identity}
Let $p=2r$ with an integer $r\ge1$, and let $s$ be a $1$-periodic $C^{p-1}$ function
whose polynomial pieces on the knot intervals have degree at most
$p$. Write $\kappa_i$ for the constant value of $s^{(p)}$ on
$(u_i,u_{i+1})$, and define
\begin{equation*}
  c_i=\kappa_i-\kappa_{i-1}\quad(0\le i\le n),
  \qquad \kappa_{-1}=\kappa_n.
  \label{eq:jumps}
\end{equation*}
Then
\begin{equation*}
  \sum_{i=0}^n c_i s(u_i)=0.
  \label{eq:jump-identity}
\end{equation*}
\end{lemma}

\begin{proof}[Proof of Lemma~\ref{lem:jump-identity}]
Integration over the knot intervals and periodic reindexing give
\begin{align*}
  \sum_{i=0}^n c_i s(u_i)
  &=-\sum_{i=0}^n\kappa_i\bigl(s(u_{i+1})-s(u_i)\bigr)=-\int_0^1s'(u)s^{(2r)}(u)\,du.
\end{align*}
The derivatives through order $p-1$ are absolutely continuous and
periodic. After $r-1$ integrations by parts, the last expression is
\begin{equation*}
  (-1)^r\int_0^1s^{(r)}(u)s^{(r+1)}(u)\,du
  =\frac{(-1)^r}{2}
    \left[\bigl(s^{(r)}(u)\bigr)^2\right]_0^1=0.
  \label{eq:periodic-integral}
\end{equation*}
All boundary terms vanish by periodicity. When $r=1$, no integration
by parts is needed. Values of the highest derivative at the knots
do not affect these integrals.
\end{proof}

\subsection{Proof of Theorem~\ref{thm:main}}

\begin{proof}
We first consider any $1$-periodic $C^{p-1}$ function $s$ whose
polynomial pieces on $[u_i,u_{i+1}]$ have degree at most $p$ and
which satisfies
\begin{equation}
  s(t_i)=0\quad(0\le i\le n).
  \label{eq:homogeneous}
\end{equation}
Periodicity also gives $s(1)=0$. We prove that these conditions
force $s\equiv0$. Suppose otherwise.

The spline cannot vanish on an open interval. If it did, one
polynomial piece would vanish identically. Continuity through order
$p-1$ at the next knot $u_i$ would force the following piece to
have the form $a(u-u_i)^p$. Each interval $(u_i,u_{i+1}]$ contains
an interpolation zero different from $u_i$: it contains $t_{i+1}$
in its interior for $0\le i<n$, and its right endpoint is the zero
$1$ for $i=n$. Substitution gives $a=0$. Repeating this argument
around the period would give $s\equiv0$, a contradiction.

Consequently, $s$ has finitely many zeros in a period, and any set
of them is separated. Lemma~\ref{lem:zeros} shows that its zeros are
precisely $t_0,\ldots,t_n$ modulo $1$. Moreover, they are simple.
Indeed, in each interval between successive parameters, choose a
point where $|s|$ attains its positive maximum. These points give
$n+1$ separated zeros of $s'$, as in the lemma's proof.
If $s'(t_j)=0$, inserting $t_j$ between the adjacent chosen points
would give one more separated zero of $s'$: on each of the
two new intervals, $s$ changes between zero and a nonzero value.
This contradicts Lemma~\ref{lem:zeros} applied to the degree-$(p-1)$
spline $s'$. Hence $s'(t_i)\ne0$ for every $i$.

The sign of $s$ changes at each $t_i$ and nowhere else. A periodic
function cannot have an odd number of such changes, so $n+1$ must
be even. Choose one point in each $(t_i,t_{i+1})$; the corresponding
values of $s$ alternate in sign cyclically.

We next pass this alternation to the highest derivative. For a
periodic, absolutely continuous, piecewise polynomial function $f$
with alternating nonzero values at cyclically ordered points
$x_0,\ldots,x_n$, put $x_{n+1}=x_0+1$. Then
\begin{equation}
  \int_{x_i}^{x_{i+1}} f'(u)\,du=f(x_{i+1})-f(x_i)
  \label{eq:alternation}
\end{equation}
has the sign of $f(x_{i+1})$. Thus each interval
$(x_i,x_{i+1})$ contains a point, away from the knots, where $f'$
has that sign. These derivative values again alternate cyclically.
Apply this argument to $s,s',\ldots,s^{(p-1)}$, all of which are
absolutely continuous and periodic. It follows that $s^{(p)}$ has
$n+1$ cyclically alternating nonzero values outside the knots.

Let $\kappa_i$ be the constant value of $s^{(p)}$ on
$(u_i,u_{i+1})$. Each knot interval contains at most one of the
selected points: all selected points in an interval are consecutive
in cyclic order, whereas a constant function cannot alternate in
sign there. There are $n+1$ points and $n+1$ intervals, so each
interval contains exactly one. Hence the $\kappa_i$ are nonzero
and alternate in sign. The jumps $c_i$ defined in
\eqref{eq:jumps} have the signs of $\kappa_i$ and therefore also
alternate.
By \eqref{eq:interlacing}, the nonzero values
$s(u_1),\ldots,s(u_n)$ alternate in sign as well. All products
$c_i s(u_i)$ for $1\le i\le n$ consequently have the same
nonzero sign, while $c_0s(u_0)=0$. Thus
\begin{equation*}
  \sum_{i=0}^n c_i s(u_i)\ne0.
  \label{eq:nonzero-sum}
\end{equation*}

This contradicts Lemma~\ref{lem:jump-identity}. Thus every spline
satisfying the homogeneous conditions \eqref{eq:homogeneous}
is identically zero.

Now take $\mathbf a=(a_0,\ldots,a_n)^{\mathsf T}\in\R^{n+1}$
with $\Nper\mathbf a=0$. The spline
$s=\sum_{j=0}^n a_j\mathcal N_{j,p}$ satisfies
\eqref{eq:homogeneous}, so $s\equiv0$. By the linear independence
established using \eqref{eq:periodic-independence}, all $a_j$ vanish.
The square matrix $\Nper$ is therefore nonsingular. Solving its
system in each coordinate proves existence and uniqueness of the
interpolating curve.
\end{proof}

\section{A counterexample to the periodic window condition}
\label{app:window-counterexample}

We first recall the condition as stated by Tang \cite{tang2023control}
(Theorems~2 and~3 therein), then give a uniform counterexample.

Let $p$ be the degree, let $0=u_0<u_1<\cdots<u_{n+1}=1$ be the simple
domain knots extended periodically by $u_{j+n+1}=u_j+1$, and let
$0=t_0<t_1<\cdots<t_n<1$ be the interpolation parameters. With
$k=\lceil p/2\rceil$, in centred indexing the periodic window condition
reads
\begin{equation}
  u_{j-k}<t_j<u_{j+k}\quad(p\text{ odd}),
  \qquad
  u_{j-k}<t_j<u_{j+k+1}\quad(p\text{ even}),
  \qquad j=0,\ldots,n,
  \label{eq:window}
\end{equation}
and is equivalent to positivity of every diagonal entry of the square
periodic collocation matrix.

The following uniform configuration shows that \eqref{eq:window} does
not imply nonsingularity at even cardinality. Take degree $p=1$,
$n+1=12$, uniform domain knots
$u_j=j/(n+1)$ and collocation parameters $t_j=(j+\tfrac12)/(n+1)$.
Then each row of the periodic collocation matrix is
$(\ldots,0,\tfrac12,\tfrac12,0,\ldots)$, and every diagonal entry
equals $\tfrac12>0$, so the window condition \eqref{eq:window} holds.
The alternating vector $v=(1,-1,\ldots,1,-1)^{\mathsf T}$ closes up
around the circle because $n+1$ is even, and every row annihilates it,
so the matrix is singular.

The same mechanism occurs at higher degrees. For $n+1=12$ and the same
uniform knots, the parameters $t_j=(j+\tfrac12)/(n+1)$ at degree $p=3$
and $t_j=j/(n+1)$ at degree $p=2$ both satisfy the window condition, 
yet the corresponding periodic collocation matrices are singular; 
the second case shows that the failure already occurs at an even degree.

\section{A counterexample to the diagonal-dominance claim for the shifting method}
\label{app:park-dominance}

Park \cite{park2001choosing} states that for the shifting method simple
pivoting makes the periodic collocation matrix strictly diagonally
dominant. Because every row of this matrix is nonnegative and sums to one,
strict diagonal dominance is equivalent to the existence of a permutation
that puts an entry larger than $1/2$ on the diagonal of every row. The
following shifted quadratic has a row without such an entry.

Let $p=2$ and $n+1=12$ (so $n=11$), and work on $[0,1)$. Take the
parameters
\begin{equation*}
  t_0,\ldots,t_{11}
  =\frac{1}{31}\,(0,2,4,13,15,17,19,21,23,25,27,29),
\end{equation*}
so that $d_n=1-t_{11}=\tfrac{2}{31}$. The shifting method of
Appendix~\ref{app:compared} replaces them by the shifted parameters
$\bar t_i=t_i+d_n/2$,
\begin{equation*}
  \bar t_0,\ldots,\bar t_{11}
  =\frac{1}{31}\,(1,3,5,14,16,18,20,22,24,26,28,30),
\end{equation*}
and sets each interior domain knot to the midpoint of two consecutive
shifted parameters,
\begin{equation*}
  u_0,\ldots,u_{11}
  =\Bigl(0,\ \frac{2}{31},\ \frac{4}{31},\ \frac{19}{62},\
  \frac{15}{31},\ \frac{17}{31},\ \frac{19}{31},\ \frac{21}{31},\
  \frac{23}{31},\ \frac{25}{31},\ \frac{27}{31},\ \frac{29}{31}\Bigr),
\end{equation*}
with $u_{12}=1$, so that $u_i=(\bar t_{i-1}+\bar t_i)/2$ for
$1\le i\le 11$ and $u_0=0$. At the shifted parameter
$\bar t_2=\tfrac5{31}$ the corresponding row of the interpolation matrix
\eqref{eq:matrix} has three nonzero entries,
\begin{equation*}
  \Bigl(\frac{27}{55},\ \frac{298}{605},\ \frac{2}{121}\Bigr)
  =\frac{1}{605}\,(297,298,10),
  \label{eq:park-row}
\end{equation*}
all strictly smaller than $1/2$. Since the row sum is one, this row has no
entry that can be brought onto the diagonal so that the diagonal entry
exceeds the sum of the remaining entries; hence no row or column
permutation makes the matrix strictly diagonally dominant. The matrix is
nevertheless nonsingular, with $\kappa_2\approx3.09$; the failure is
therefore one of the dominance argument, not of solvability.

\section{Compared methods}
\label{app:compared}

Two deterministic even-degree constructions serve as baselines
\cite{park2001choosing}: the shifting method and the energy method. The
\emph{shifting} method is the standard
deterministic construction for even degree. With $d_n=1-t_n$ it shifts the parameters to
$\bar t_i=t_i+d_n/2$ and places each interior knot at the midpoint of two
consecutive shifted parameters,
\begin{equation}
  u_0=0,\qquad u_i=\frac{\bar t_{i-1}+\bar t_i}{2}\quad(1\le i\le n),
  \qquad u_{n+1}=1,
  \label{eq:shifting}
\end{equation}
so that every shifted parameter is exactly centered in its knot interval.

The \emph{energy} method sets $\hat n=n+1$, so that it uses one extra
control point, places the centered knots
\begin{equation}
  u_0=0,\qquad u_i=\frac{t_{i-1}+t_i}{2}\quad(1\le i\le \hat n),
  \qquad u_{\hat n+1}=1,
  \label{eq:energy-knots}
\end{equation}
with $t_{n+1}=1$, and minimizes the same energy \eqref{eq:fairness}
directly under the $n+1$ interpolation constraints. The constrained
minimum is characterized by the KKT system
\begin{equation}
  \begin{pmatrix}
    \mathbf K & \mathbf N^{\mathsf T}\\[2pt]
    \mathbf N & \mathbf 0
  \end{pmatrix}
  \begin{pmatrix}
    \mathbf P\\[2pt] \boldsymbol\lambda
  \end{pmatrix}
  =
  \begin{pmatrix}
    \mathbf 0\\[2pt] \mathbf Q
  \end{pmatrix},
  \label{eq:kkt}
\end{equation}
where $\mathbf N$ is the $(n+1)\times(\hat n+1)$ periodic collocation
matrix, $\mathbf K$ is the stiffness matrix \eqref{eq:stiffness} built from
the periodic B-spline basis, and $\boldsymbol\lambda$ collects the Lagrange
multipliers. The interpolant is obtained by solving this linear system.

\section{More examples}
\label{app:shapes}

\begin{figure}[H]
  \centering
  \includegraphics[width=1.0\textwidth]{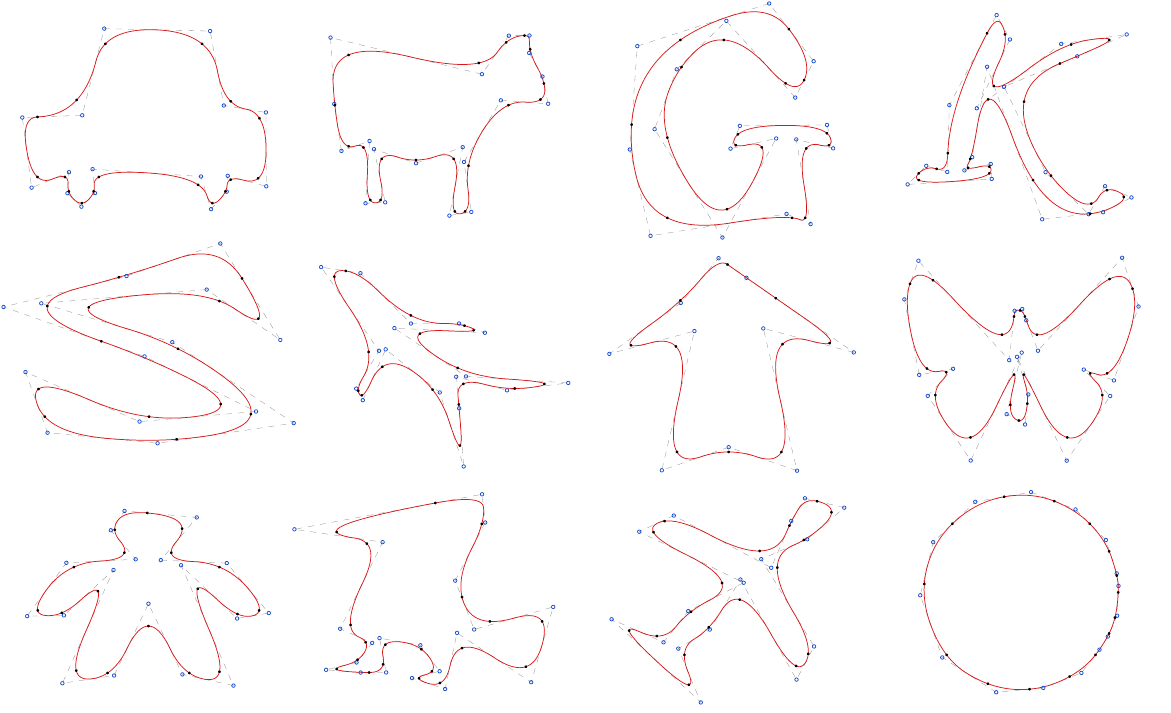}
  \caption{Additional test shapes, reconstructed at degree $p=2$
  with centripetal parametrization; the interlacing alternates between Opt-R
  and Opt-L.}
  \label{fig:remaining}
\end{figure}

Figure~\ref{fig:remaining} shows additional closed polygonal data sets,
reconstructed at degree $p=2$ under centripetal parametrization. Red
curves are the interpolants, blue circles and grey dashed segments are the
control points and the control polygon, and black points are the data points.
The interlacing alternates between Opt-R and Opt-L. The optimized interlacing
works well on all of them, and every reconstructed curve is visually pleasing.

\bibliographystyle{unsrtnat}
\bibliography{reference}

\end{document}